\documentclass[12pt,onecolumn,draftcls]{IEEEtran}
\usepackage[T1]{fontenc}
\usepackage{url}
\usepackage{amsmath}
\usepackage{amsthm}
\usepackage{amssymb}
\usepackage{graphicx}
\usepackage[unicode=true,
 bookmarks=true,bookmarksnumbered=true,bookmarksopen=true,bookmarksopenlevel=1,
 breaklinks=false,pdfborder={0 0 0},pdfborderstyle={},backref=false,colorlinks=false]
 {hyperref}
\hypersetup{pdftitle={Your Title},
 pdfauthor={Your Name},
 pdfpagelayout=OneColumn, pdfnewwindow=true, pdfstartview=XYZ, plainpages=false}

\makeatletter

\providecommand{\tabularnewline}{\\}

\theoremstyle{plain}
\newtheorem{thm}{\protect\theoremname}
\theoremstyle{remark}
\newtheorem{rem}[thm]{\protect\remarkname}
\theoremstyle{definition}
\newtheorem{defn}[thm]{\protect\definitionname}

\usepackage[caption=false,font=footnotesize]{subfig}
\usepackage{multicol}
\usepackage{graphicx}

\usepackage[noadjust]{cite}

\@ifundefined{showcaptionsetup}{}{%
 \PassOptionsToPackage{caption=false}{subfig}}
\usepackage{subfig}
\makeatother

\providecommand{\definitionname}{Definition}
\providecommand{\remarkname}{Remark}
\providecommand{\theoremname}{Theorem}

\begin{document}

\title{Fine-Grained Reliability for V2V Communications around Suburban and
Urban Intersections}

\author{Mouhamed~Abdulla,~\IEEEmembership{Member,~IEEE,} and~Henk~Wymeersch,~\IEEEmembership{Member,~IEEE}
\thanks{The authors are with the Division of Communication and Antenna Systems,
Dept. of Electrical Engineering, Chalmers University of Technology,
G�teborg, Sweden, e-mails: \{mouhamed,henkw\}@chalmers.se. \protect \\
This research work is supported, in part, by the European Commission
under the Marie Sk\l odowska-Curie Individual Fellowship (H2020-MSCA-IF-2014),
EU-MARSS-5G project, Grant No. 659933; Ericsson Research Foundation,
Grant No. FOSTIFT-16:043-17:054; the VINNOVA COPPLAR project, funded
under Strategic Vehicle Research and Innovation Grant No. 2015-04849;
and the EU-H2020 HIGHTS project (High Precision Positioning for Cooperative
ITS Applications), Grant No. MG-3.5a-2014-636537.\protect \\
This draft is also available on arXiv, URL: \protect\url{https://arxiv.org/abs/1706.10011}.}}
\maketitle
\begin{abstract}
Safe transportation is a key use-case of the 5G/LTE Rel.15+ communications,
where an end-to-end reliability of 0.99999 is expected for a vehicle-to-vehicle
V2V transmission distance of 100\textendash 200 m. Since communications
reliability is related to road-safety, it is crucial to verify the
fulfillment of the performance, especially for accident-prone areas
such as intersections. We derive closed-form expressions for the V2V
transmission reliability near suburban corners and urban intersections
over finite interference regions. The analysis is based on plausible
street configurations, traffic scenarios, and empirically-supported
channel propagation. We show the means by which the performance metric
can serve as a preliminary design tool to meet a target reliability.
We then apply meta distribution concepts to provide a careful dissection
of V2V communications reliability. Contrary to existing work on infinite
roads, when we consider finite road segments for practical deployment,
fine-grained reliability per realization exhibits bimodal behavior.
Either performance for a certain vehicular traffic scenario is very
reliable or extremely unreliable, but nowhere in relatively proximity
to the average performance. In other words, standard SINR-based average
performance metrics are analytically accurate but can be insufficient
from a practical viewpoint. Investigating other safety-critical point
process networks at the meta distribution-level may reveal similar
discrepancies.
\end{abstract}

\section{\label{Sec1: Introduction}Introduction }

\IEEEPARstart{T}{he} United Nations World Health Organization estimates
that 1.25\,million people are fatally injured every year due to road-traffic
accidents \cite{UN_WHO_1.2Maccidents}. Studies conducted by the US
Department of Transportation suggests that 2\% of crashes are due
to vehicle components failure or degradation; another 2\% are attributed
to the environment and weather conditions, while a staggering 94\%
is tied to human choice or error, such as careless driving, speeding,
and driving under the influence \cite{DOT_2015_94perc}. This alarming
reality suggests that human cognition is insufficient in its capacity
to maneuver in an intricate transportation system with high reliability.
Reliance on augmented technology is thus needed to compensate for
the limitations of human drivers. Eventually, highly automated vehicles
(HAV) will operate in self-driving capability through 360-degree awareness
of their surroundings via artificial intelligence, machine learning
and cooperative communications.

Present manifestations of autonomous vehicles  rely predominantly
on radars, cameras, LiDARs, ultrasonic sensors, GPS, and cloud-based
vehicle-to-network 3D digital mapping. Sensor-based autonomous vehicles
are, however, constrained by line-of-sight, and their effectiveness
is significantly influenced by weather conditions, such as fog, sunbeams,
heavy rain and snow. Meanwhile, vehicle-to-vehicle (V2V) communications
is the \textit{only} HAV-related technology that has the capacity
to \textit{see around corners} in the presence of urban structures.
Moreover, V2V communications is more reliable than sensors in harsh
weather conditions. Although current efforts are still at the research
phase, dynamically formed vehicular ad hoc network (VANET) will inevitably
be included within the HAV ecosystem. Evidently, the addition of V2V
capability as a complementary technology and as a supplemental source
of system redundancy is expected to improve the reliability of machine-driven
units and, as a consequence, further shrink the likelihood of accidents
by HAV-systems \cite{DOT_Automated2016,Moe_McGill16}. And this is
not surprising, given that standalone V2V technology is estimated
to prevent up to 35\% of serious road accidents \cite{EU_V2V_35perc}.

\begin{figure}[t]
\begin{centering}
\includegraphics[width=0.7\columnwidth]{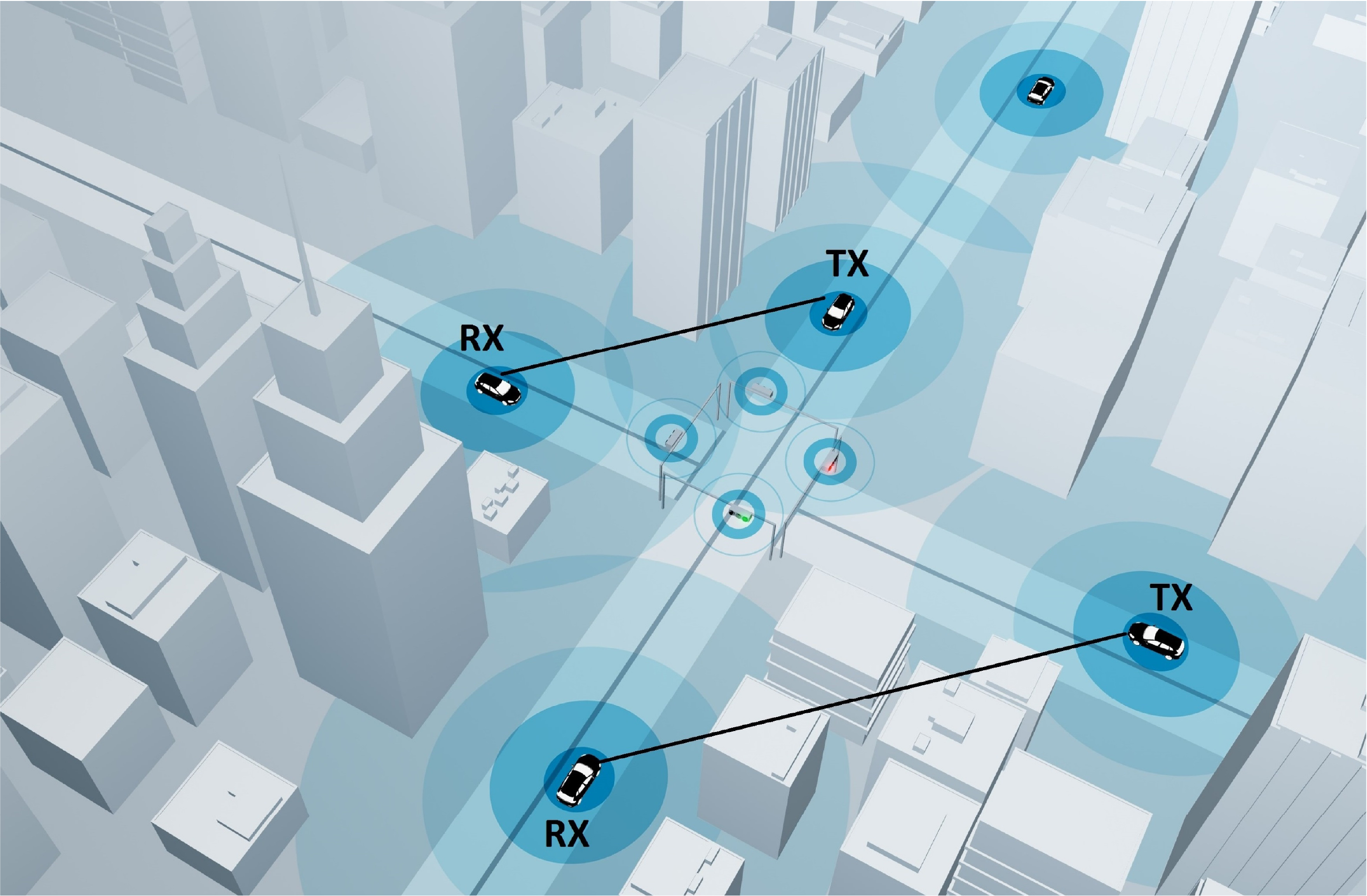}
\par\end{centering}
\centering{}\caption{Intelligent V2V wireless communications of safety-critical data by
a pair of vehicles maneuvering around a built-up blind urban intersection,
whereby a transmitter (TX) sends data packets to a receiver (RX),
in the presence of simultaneous interferers (source: image adapted
from and used with permission from Volvo Car Group, Public Affairs,
SE-405 31, G�teborg, Sweden). \label{fig1: Intersection Pic}}
\end{figure}

There are various V2V use-cases that require careful investigation
and analysis for the purpose of enhancing road-safety and traffic
efficiency. Data assessed between 2010 to 2015 suggest that the nearly
half of all vehicular accidents occur at intersections \cite{DOT_2015_47.24perc}.
Despite vehicular high-densification and greater proliferation of
accident avoidance technology, it is surprising to note that the rate
of intersection accidents remained consistent year after year. These
numbers reveal that intersection-related accidents are high consequence
events that occur with a very high probability. Thus, carefully studying
the reliability of V2V communications around intersections, and in
particular blind urban junctions where the loss due RF propagation
is of major concern (see Fig.~\ref{fig1: Intersection Pic}), will
help us assess the feasibility of seeing non-line-of-sight vehicles
near corners.

\subsection{Enabling Technology}

To ensure effective scalability and interoperability among radio units
mounted on vehicles, standardization is of utmost importance. Today,
there are two competing standards: (i) IEEE 802.11p, commonly known
as direct short range communication (DSRC)\footnote{DSRC for dynamic vehicles relatively resembles the popular Wi-Fi (IEEE
802.11) standard utilized for quasi-stationary wireless connectivity.} \cite{IEEE802.11P}; and (ii) cellular-V2X (C-V2X)\footnote{C-V2X also supports direct VANET communication (i.e., similar to DSRC)
without necessarily relying on network involvement for scheduling.} connectivity supported by 4G/LTE Rel.14+ \cite{LTERel.14_Jun2015,LTERel.14_Dec2015,5GAA-LTE}.
The competition among these two standards is still fluid and continuously
evolving. Granted, the locally-based DSRC standard is defined and
ready for utilization, whereas the network-based C-V2X is still under
development for 5G/LTE Rel.15+ operation aimed around 2020. Meanwhile,
HAV-systems will rely on machine learning capability, where hours
of driving experience will serve as an acquisition mechanism of observing
and learning from experience. Newly acquired knowledge will then be
shared wirelessly to the cloud and to other units using V2X capability
at high data rates. Other vehicles will immediately upgrade their
knowledge-based system regarding the transportation status and aggregate
experience of certain geographical regions without the need for further
acquisition time.

Communicating real-time big data generated by HAVs will require the
potential of 5G broadband connectivity where a peak data rate of 10
Gbps is projected \cite{EU_5G_10Gbps}. It is reassuring to know that
such multi-Gbps transmission is envisioned and possible using technologies
such as mmWave communications \cite{mmWaveV2x_Heath_ComSocMag16,mmWaveV2x_China_ComSocMag17}.
Moreover, C-V2X is expected to deliver a larger communications range;
by some estimates twice the distance of DSRC for a fixed reliability
value. Further, the reliability under 5G performs significantly better
than DSRC \cite{EricssonV2x_Jan17}. Also, greater mobility of up
to 500\,km/hr is supported. Overall, the reliance on C-V2X operating
with 5G seems promising and quite compatible with the requirements
of HAV-systems.

\subsection{Problem Statement and Contribution}

Ultimately, we need to develop analytical expressions that will serve
as a mechanism to quantify the extent of reliability for a certain
V2V communication link and eventually aid in identifying the contribution
of relevant parameters for the purpose of network design. Analytical
modeling based on point processes is well suited to study such problems
where techniques from stochastic geometry have been applied to vehicular
communications (e.g. \cite{Bartek09_VANET,Bartek12_VANET,MoeWin2013,Erik_GC15,Hesham2016_Highway,Haenggi2016_DSRC,Japan16_VTCStochasticGeometry,Erik_arXiv16,Moe_GC16}).
As for intersections, they were explicitly considered in \cite{Erik_GC15},
though only for suburban/rural scenarios over infinitely long roads.
For the analytical expressions to have practical real-world relevance,
they must build on plausible VANET scenarios coupled with channel
models validated by extensive measurement campaigns \cite{Tufvesson2011IEEEMag,Tufvesson2011TVT}.
The above-mentioned works in stochastic geometry allow the evaluation
of the \emph{average reliability}, obtained by averaging over different
fading realizations and node placements. This average reliability
may obscure the performance for specific node configuration \cite{Haenggi2016_MetaDistribution},
referred to as the \emph{fine-grained reliability}.

In this paper, we perform a study of average and fine-grained reliability,
dedicated to urban intersections, which have particular propagation
characteristics \cite{Tufvesson_4IntersectionCases_2010,UrbanIntersection_OtherGerman11,BMW2011_Virtual_11p,Tfvesson-Abbas2013_Sweden},
complementing and generalizing the results in \cite{Erik_GC15} and
\cite{Moe_GC16}. The key contributions of this paper are summarized
as follows:
\begin{enumerate}
\item We evaluate the transmission reliability as a function of system parameters,
vehicular traffic, infrastructure geometry, and empirically-validated
V2V channel propagation models geared for suburban and urban intersections.
The analysis is generic and allows for closed-form expressions over
finite road segments of practical significance for real-world deployment.
\item While considering a large number of design parameters embedded within
the derived reliability metrics, we demonstrate the approach for re-configuring
the VANET in order to ascertain a target reliability.
\item Using extensive Monte Carlo simulation techniques, we show the fine-grained
V2V reliability obtained through the meta distribution. Although analytically
accurate, the discrepancy between\emph{ }\textit{\emph{average reliability}}
and \textit{\emph{fine-grained reliability}}\emph{ }is substantial
from a practical standpoint. The most striking observation is that
average reliability provides an oversimplified distortion of the actual
communication performance incurred by each vehicular traffic realization
under the specified VANET scenario.
\end{enumerate}

\subsection{Organization}

The rest of the paper is organized as follows. In Section \ref{Sec2: System Model},
we outline the considered VANET network traffic and geometry and describe
the specialized intersection-based channel models for V2V communications
under line-of-sight (LOS), weak-line-of-sight (WLOS), and non-line-of-sight
(NLOS) scenarios. Then, in Section \ref{Sec3: Generalized Transmission Reliability},
we dissect the stipulated requirements for reliability in 5G communications;
we provide an analytical definition for reliability and relate it
to its average; we also explain the notion of fine-grained reliability.
Closed-form performance measures for different channel environments
are subsequently derived in Section \ref{Sec4: Exact Reliability for Different Channels}
and applied to  network design in order to meet a preset target performance.
Furthermore, in Section \ref{Sec6: Simulations and Discussion}, we
display results from fine-grained reliability based on extensive Monte
Carlo simulations of the meta distribution, and we discuss its unexpected
relation to standard reliability assessment obtained through averages.
Finally, we conclude the paper with Section \ref{Sec7: Conclusion}.

\begin{figure*}[t]
\begin{centering}
\subfloat[suburban channel: LOS/WLOS\label{fig2a: Suburban Channel (LOS/WLOS)}]{\begin{centering}
\includegraphics[width=0.24\columnwidth]{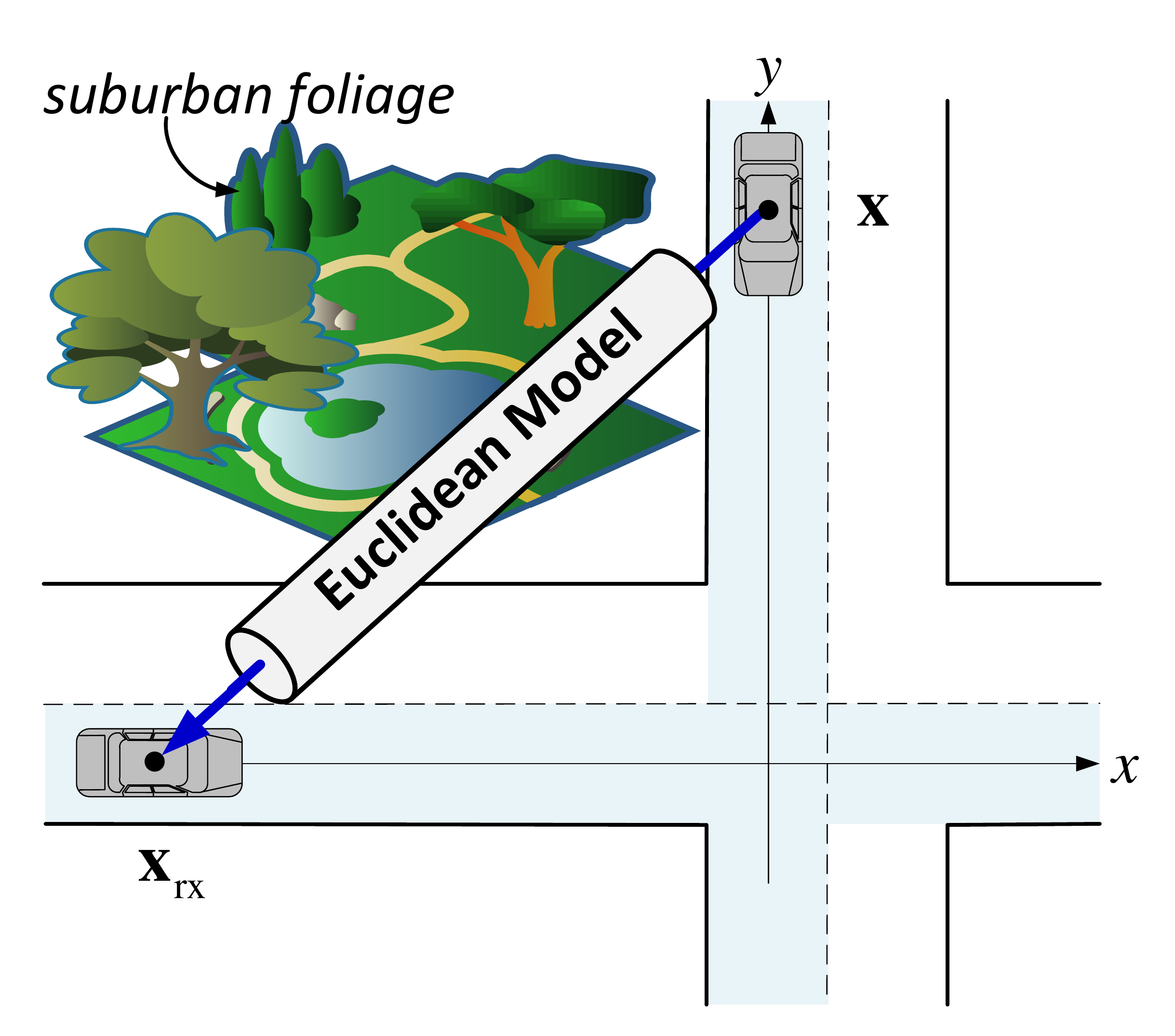}
\par\end{centering}
} \subfloat[urban channel: NLOS\label{fig2b: Urban Channel (NLOS)}]{\begin{centering}
\includegraphics[width=0.24\columnwidth]{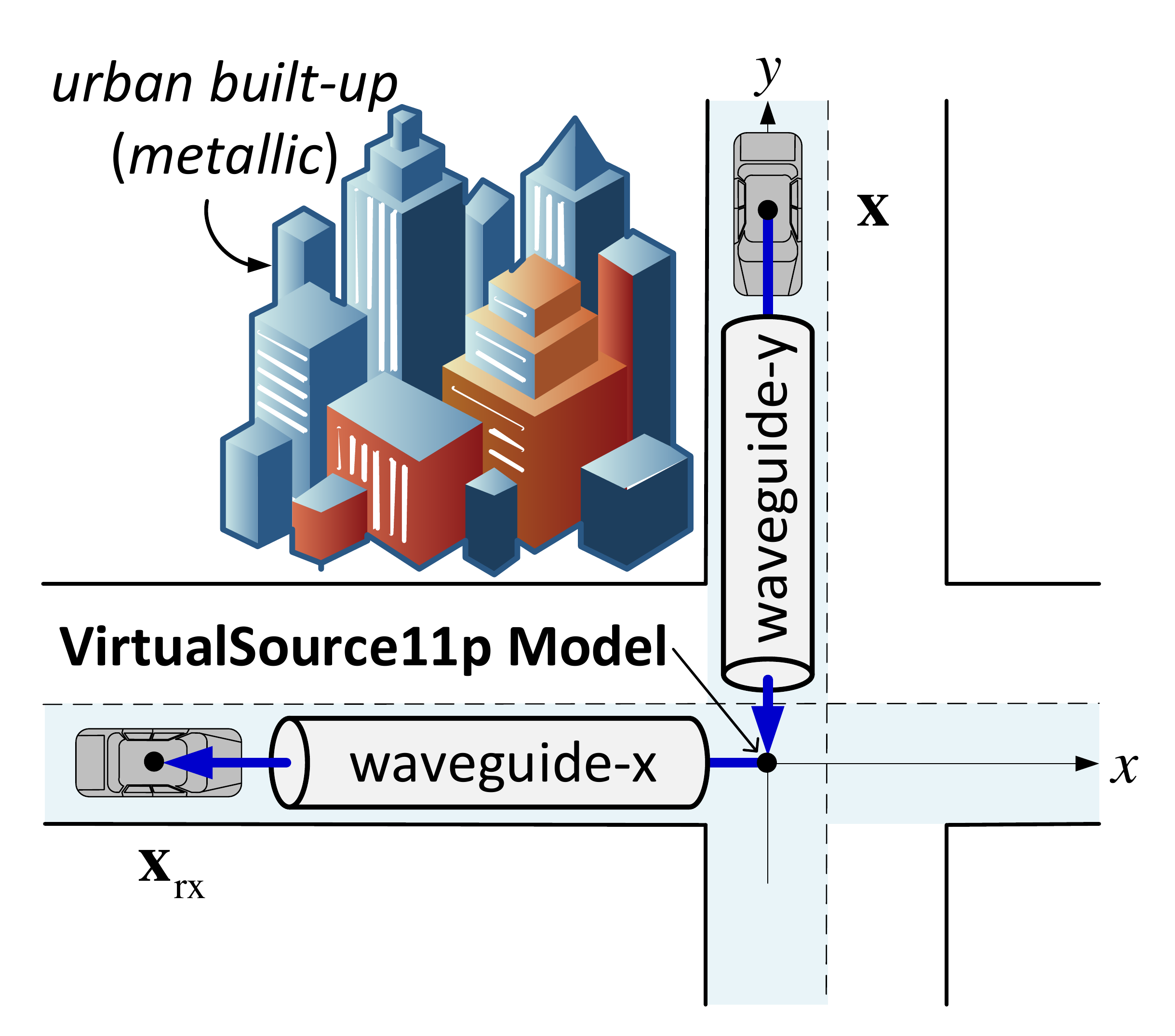}
\par\end{centering}
} \subfloat[urban channel: WLOS\label{fig2c: Urban Channel (WLOS)}]{\begin{centering}
\includegraphics[width=0.24\columnwidth]{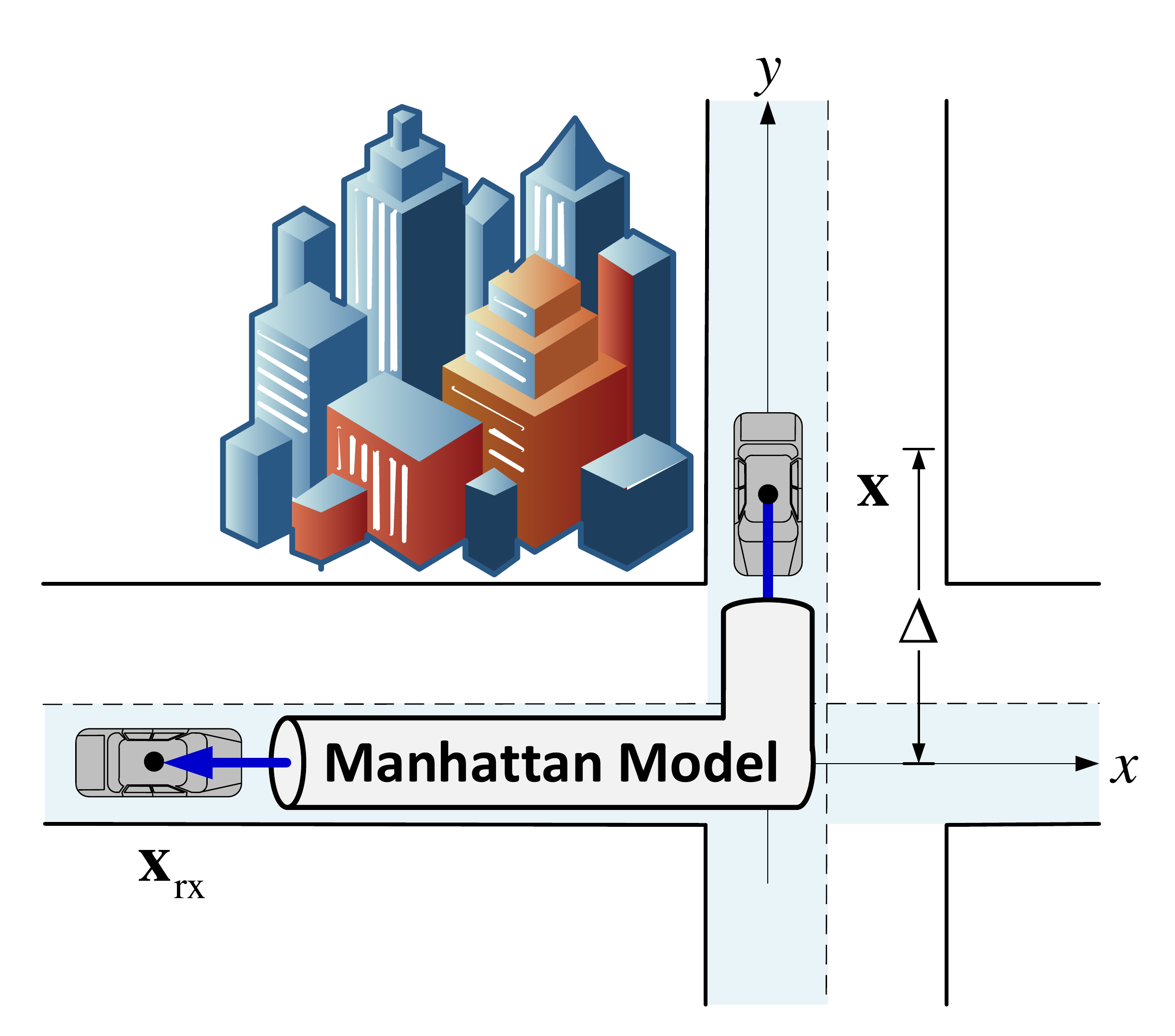}
\par\end{centering}
} \subfloat[urban channel: LOS\label{fig2d: Urban Channel (LOS)}]{\begin{centering}
\includegraphics[width=0.24\columnwidth]{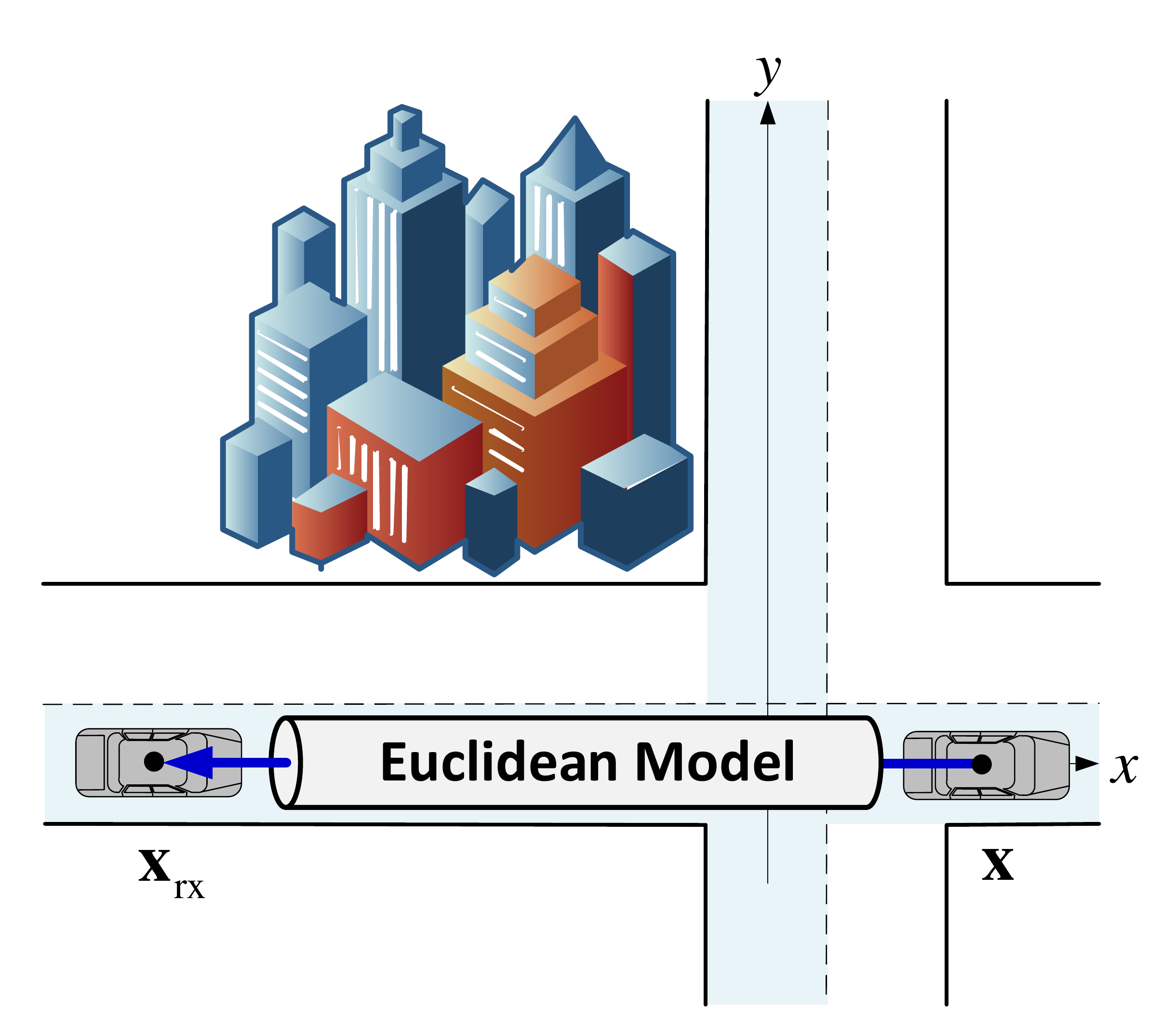}
\par\end{centering}
}
\par\end{centering}
\centering{}\caption{Propagation models for different V2V channel environments as a function
of vehicular positions around suburban corners and blind urban intersections.\label{fig2: Channel Models}}
\end{figure*}

\section{\label{Sec2: System Model}System Model}

\subsection{\label{Sec2.1: Network Model}Network Model}

The analysis is based on VANET formed near suburban corners and blind
urban intersections, where the network traffic and geometry is described
as follows. The transmitter (TX) can be located anywhere on the horizontal
or vertical road, and the receiver (RX), which, without loss of generality,
is confined to the horizontal road. Hence, $\mathbf{x}_{\text{\ensuremath{\mathrm{tx}}}}=x_{\mathrm{tx}}\mathbf{e}_{\mathrm{x}}+y_{\mathrm{tx}}\mathbf{e}_{\mathrm{y}}$
and $\mathbf{x}_{\mathrm{rx}}=x_{\mathrm{rx}}\mathbf{e}_{\mathrm{x}}$,
where $x_{\mathrm{tx}},x_{\mathrm{rx}},y_{\mathrm{tx}}\in\mathbb{R}$,
such that $x_{\mathrm{tx}}y_{\mathrm{tx}}=0$, where $\mathbf{e}_{\mathrm{x}}=[1\,0]^{\mathrm{T}}$,
$\mathbf{e}_{\mathrm{y}}=[0\,1]^{\mathrm{T}}$. Other vehicles are
randomly positioned on both horizontal and vertical roads and follow
a homogeneous Poisson point process (H-PPP) over bounded sets $\mathcal{B}_{\mathrm{x}}=\left\{ x\in\mathbb{R}\bigl|\left|x\right|\leq R_{\mathrm{x}}\right\} $
and $\mathcal{B}_{\mathrm{y}}=\left\{ y\in\mathbb{R}\bigl|\left|y\right|\leq R_{\mathrm{y}}\right\} $,
such that $R_{\mathrm{x}}>0$ and $R_{\mathrm{y}}>0$ are road segments
of the intersection region, and the vehicular traffic intensities
are respectively given by $\lambda_{\mathrm{x}}$ and $\lambda_{\mathrm{y}}$.
Moreover, interfering vehicles follow an Aloha MAC protocol\footnote{Resource selection for DSRC is based on CSMA with collision avoidance;
and C-V2X defined by 3GPP-PC5 interface relies on semi-persistent
transmission with relative energy-based selection. Nonetheless, for
the purpose of preliminary analysis, we here only consider the Aloha
MAC protocol.} and can transmit independently with a probability $p_{\mathrm{I}}\!\in\!\left[0,1\right]$.
In the rest of the paper, the following shorthand notations are accordingly
used to refer to the geometry of interfering vehicles on each road,
modeled by thinned H-PPPs
\begin{align}
\Phi_{\mathrm{x}} & =\left\{ \mathbf{x}_{i}\right\} _{i=1,2,\ldots,n}\ \in\mathbb{R}^{n}\:\sim\textrm{PPP}\left(p_{\mathrm{I}}\lambda_{\mathrm{x}},\mathcal{B}_{\mathrm{x}}\right)\label{eq: PPP-X}\\
\Phi_{\mathrm{y}} & =\left\{ \mathbf{x}_{j}\right\} _{j=1,2,\ldots,m}\in\mathbb{R}^{m}\sim\textrm{PPP}\left(p_{\mathrm{I}}\lambda_{\mathrm{y}},\mathcal{B}_{\mathrm{y}}\right),\label{eq: PPP-Y}
\end{align}
such that $n$ and $m$ are random Poisson distributed integers with
mean $p_{\mathrm{I}}\lambda\left|\mathcal{B}\right|$, where $\left|\mathcal{B}\right|$
is the Lebesgue measure of bounded set $\mathcal{B}$. All vehicles,
including TX, broadcast with the same power level $P_{\circ}$. The
receiver signal-to-interference-plus-noise-ratio ($\mathsf{SINR}$)
threshold for reliable detection is set to $\beta$, in the presence
of additive white Gaussian noise (AWGN) with power $N_{\circ}$. The
$\mathsf{SINR}$ depends on the propagation channel, described next.

\subsection{\label{Sec2.2: Channel Models}Channel Models}

The detected power observed at the RX from an active transmitting
vehicle at location $\mathbf{x}$ is modeled by $P_{\mathrm{rx}}\left(\mathbf{x},\mathbf{x}_{\mathrm{rx}}\right)=P_{\circ}\ell_{\mathrm{ch}}\left(\mathbf{x},\mathbf{x}_{\mathrm{rx}}\right)$,
which depends on transmit power $P_{\circ}$ and channel losses $\ell_{\mathrm{ch}}\left(\mathbf{x},\mathbf{x}_{\mathrm{rx}}\right)$.
The channel losses consist of three components: average path loss
$\ell_{\mathrm{pl}}\left(\mathbf{x},\mathbf{x}_{\mathrm{rx}}\right)$
that captures the propagation losses, shadow fading $\ell_{\mathrm{s}}\left(\mathbf{x},\mathbf{x}_{\mathrm{rx}}\right)$
that captures the effects of random channel obstacles, and random
small-scale fading $\ell_{\mathrm{f}}\left(\mathbf{x}\right)$ that
captures the non-coherent addition of signal components. For the purpose
of tractability, we implicitly consider shadow fading to be inherent
within the H-PPP, and thus regard $\ell_{\mathrm{ch}}\left(\mathbf{x},\mathbf{x}_{\mathrm{rx}}\right)\simeq\ell_{\mathrm{pl}}\left(\mathbf{x},\mathbf{x}_{\mathrm{rx}}\right)\ell_{\mathrm{f}}\left(\mathbf{x}\right)$
\cite{RossEquivalence17,Bartek13_Equivalence}. We model $\ell_{\mathrm{f}}\left(\mathbf{x}\right)\sim\textrm{Exp}\left(1\right)$
as Rayleigh fading, independent with respect to $\mathbf{x}$. The
path loss $\ell_{\mathrm{pl}}\left(\mathbf{x},\mathbf{x}_{\mathrm{rx}}\right)$
for different channel environments is described below.

\subsubsection{\label{Sec2.2.1: Suburban Channel}Suburban/Rural Channel}

As shown on Fig.~\ref{fig2a: Suburban Channel (LOS/WLOS)}, the average
path loss model that is generally considered for V2V propagation adheres
to an inverse power-law \cite{Tufvesson2011TVT,Tufvesson2011IEEEMag};
thus
\begin{align}
\ell_{\mathrm{pl}}^{\mathrm{s}}\left(\mathbf{x},\mathbf{x}_{\mathrm{rx}}\right)=A_{\circ}\left\Vert \mathbf{x_{\mathrm{rx}}}-\mathbf{x}\right\Vert ^{-\alpha} & \ \ \ \ \mathbf{x}\neq\mathbf{x_{\mathrm{rx}}}\mathrm{.}\label{eq: Channel - Suburban}
\end{align}
In this expression, $\left\Vert \cdot\right\Vert $ is the $l_{2}$-norm,
$A_{\circ}$ corresponds to the LOS/WLOS path loss coefficient, which
is primarily a function of operating frequency $f_{\circ}$, path
loss exponent $\alpha\!>\!1$, reference distance $d_{\circ}$, and
antenna heights $h\left(\mathbf{x}\right)$ and $h\left(\mathbf{x}_{\mathrm{rx}}\right)$.

\subsubsection{\label{Sec2.2.2: Urban Channel}Urban Channel}

For metropolitan intersections where the concentration of high-rise
and impenetrable metallic-based buildings and structures are prevalent,
the previous model is rather unrealistic. As a consequence, a specialized
channel predictor is required. Real-world measurements of V2V communications
operating at 5.9\,GHz were conducted at different urban intersection
locations within the city of Munich, Germany. This led to the development
of the VirtualSource11p path loss model \cite{BMW2011_Virtual_11p},
where the name refers to a virtual-relay at the junction point. A
subsequent study conducted in Lund, Sweden independently validated
the accuracy of this model \cite{Tfvesson-Abbas2013_Sweden}. As a
result, it serves as inspiration to our simplified model that accounts
for all possible channel occurrences as vehicles navigate around the
intersection
\begin{align}
 & \ell_{\mathrm{pl}}^{\mathrm{u}}\left(\mathbf{x},\mathbf{x}_{\mathrm{rx}}\right)=\label{eq: Channel - Urban}\\
 & \ \ \ \ \ \ \begin{cases}
A_{\circ}^{\prime}\left(\left\Vert \mathbf{x}\right\Vert \cdot\left\Vert \mathbf{x}_{\mathrm{rx}}\right\Vert \right)^{-\alpha} & \min\left(\left\Vert \mathbf{x}\right\Vert =\left|y\right|,\left\Vert \mathbf{x}_{\mathrm{rx}}\right\Vert \right)>\triangle\\
A_{\circ}\left(\left\Vert \mathbf{x}\right\Vert +\left\Vert \mathbf{x}_{\mathrm{rx}}\right\Vert \right)^{-\alpha} & \min\left(\left\Vert \mathbf{x}\right\Vert =\left|y\right|,\left\Vert \mathbf{x}_{\mathrm{rx}}\right\Vert \right)\leq\triangle\\
A_{\circ}\left\Vert \mathbf{x}_{\mathrm{rx}}-\mathbf{x}\right\Vert ^{-\alpha} & \left\Vert \mathbf{x}\right\Vert =\left|x\right|;\ \mathbf{x}\neq\mathbf{x}_{\mathrm{rx}}\mathrm{.}
\end{cases}\nonumber 
\end{align}
As depicted in Fig.~\ref{fig2b: Urban Channel (NLOS)}, the first
case refers to the VirtualSource11p model, where $\mathbf{x}=y\mathbf{e}_{\mathrm{y}}$
is in NLOS, orthogonal to the RX. The second case refers to the Manhattan
model for WLOS when either TX/interferer or RX are close to the intersection;
this is characterized by the break-point distance $\triangle$ as
shown in Fig.~\ref{fig2c: Urban Channel (WLOS)}, typically on the
order of the road width. The third case refers to the standard Euclidean
model, where $\mathbf{x}=x\mathbf{e}_{\mathrm{x}}$ is in LOS, on
the same road as the RX as shown in Fig.~\ref{fig2d: Urban Channel (LOS)}.
Meanwhile, since the propagation under NLOS is more severe than the
cases under LOS/WLOS, then the NLOS path loss coefficient $A_{\circ}^{\prime}$
should generally satisfy the condition that $A_{\circ}^{\prime}\!<\!A_{\circ}\left(\triangle/2\right)^{\alpha}$,
where the determination of $A_{\circ}$ was described earlier.
\begin{rem}
\noindent The model in (\ref{eq: Channel - Urban}) exhibits discontinuities.
A mixture (a linear weighting) of these models can be used to avoid
these discontinuities, though this is not considered in this paper.
\end{rem}
\noindent 
\begin{rem}
\noindent We will only consider the plausible case where the region
of H-PPP interferers is greater or equal than the path loss break-point
distance, i.e., $\min\left(R_{\mathrm{x}},R_{\mathrm{y}}\right)\!\geq\!\triangle$. 
\end{rem}

\section{\label{Sec3: Generalized Transmission Reliability}Generalized Transmission
Reliability }

For road-safety purposes, a 5G reliability of $10^{-5}$ is required
up to transmission distance of $200$\,m in the suburbs, and $100$\,m
in urban settings \cite{EU_5GVision_10-5}. With limited cross-layer
optimization, this end-to-end requirement can be dissected as follows:
$10^{-2}\!\sim\!10^{-3}$ at the physical (PHY) layer; $10^{-4}$
at the MAC layer (with at most $2\!\sim\!3$ low-latency retransmissions);
and $10^{-5}$ at the network layer. Such reliability categorization
is generally consistent for semi-autonomous vehicles operating with
dual LiDAR/V2X capabilities. Full autonomy will require stringent
ultra-high reliability in the range of $10^{-7}\!\sim\!10^{-9}$.
To assess these values for different VANET scenarios, we require:
(i) a definition for reliability, and (ii) a general evaluation mechanism
for reliability.

\subsection{\label{Sec3.1: Defining Transmission Reliability}Defining Reliability}

There are multiple ways to define reliability, the most prevalent
among them are:
\begin{itemize}
\item \textit{Rate Coverage Probability}: This metric, suitable for long
packets, refers to the likelihood that the achievable rate $R_{\mathrm{ar}}=B\log_{2}\left(1+\mathsf{SINR}\right)$
(measured in bits per seconds) at the RX, where $B$ is the bandwidth
of the communication link, is larger than a threshold $r_{\mathrm{th}}$;
i.e., $\Pr\left(R_{\mathrm{ar}}\geq r_{\mathrm{th}}\right)=\Pr\left(\mathsf{SINR}\geq\beta\right)$,
such that the $\mathsf{SINR}$ threshold corresponds to $\beta\triangleq2^{r_{\mathrm{th}}/B}-1$.
The link layer outage probability is defined as: $1-\Pr\left(\mathsf{SINR}\geq\beta\right)$.
\item \textit{Maximum Coding Rate}: This metric, suitable for short packets,
is a function of the finite packet length $n_{\mathrm{p}}\simeq BT$,
where $T$ is the transmission duration; and finite packet error rate
(PER) $\epsilon_{\mathrm{p}}$. The rate is approximated by $R^{\ast}\left(n_{\mathrm{p}},\epsilon_{\mathrm{p}}\right)\simeq R_{\mathrm{ar}}-\sqrt{V/n_{\mathrm{p}}}Q^{-1}\left(\epsilon_{\mathrm{p}}\right)$,
where $Q^{-1}\left(\cdot\right)$ is the inverse of the Gaussian Q-function
and $V$ is channel dispersion \cite{GiuseppePopovski2016,Polyanskiy2010}.
In other words, to sustain a target PER over a finite packet size,
a penalty is incurred to the achievable rate.
\end{itemize}
Although the maximum coding rate is a generalization of the rate coverage
probability, analysis in conjunction with point processes is yet to
be developed. As a result, we will consider the rate coverage definition
for reliability, detailed next.

\subsection{\label{Sec3.2: Expected Reliability}Average Reliability}

Our goal is to determine the success probability $\mathcal{P}_{\mathrm{c}}\left(\beta,\mathbf{x}_{\text{\ensuremath{\mathrm{tx}}}},\mathbf{x}_{\mathrm{rx}}\right)\triangleq\Pr\left(\mathsf{SINR}\geq\beta\right)$,
where $\Pr\left(\cdot\right)$ is averaged over small-scale fading
of the wanted link $\ell_{\mathrm{f}}\left(\mathbf{x}_{\text{\ensuremath{\mathrm{tx}}}}\right)$
and point processes $\Phi_{\mathrm{x}}$ and $\Phi_{\mathrm{y}}$;
and where
\begin{align}
\mathsf{SINR} & =\frac{\ell_{\mathrm{f}}\left(\mathbf{x}_{\text{\ensuremath{\mathrm{tx}}}}\right)\ell_{\mathrm{pl}}\left(\mathbf{x}_{\text{\ensuremath{\mathrm{tx}}}},\mathbf{x}_{\mathrm{rx}}\right)}{{\displaystyle \sum_{\mathbf{x}\in\Phi_{\mathrm{x}}\cup\Phi_{\mathrm{y}}}\ell_{\mathrm{f}}\left(\mathbf{x}\right)\ell_{\mathrm{pl}}\left(\mathbf{x},\mathbf{x}_{\mathrm{rx}}\right)+\gamma_{\circ}}},\label{eq: SINR =00003D S/(I+N)}
\end{align}
in which $\gamma_{\circ}\!\!=\!\!N_{\circ}/P_{\circ}$. We introduce
a normalized aggregate interference associated with some process $\Phi$
as: $I\left(\Phi\right)\triangleq\sum_{\mathbf{x}\in\Phi}\ell_{\mathrm{f}}\left(\mathbf{x}\right)\ell_{\mathrm{pl}}\left(\mathbf{x},\mathbf{x}_{\mathrm{rx}}\right)$.
After isolating for the exponentially distributed fading of the useful
signal, and taking the expectation of the probability with respect
to interference, we obtain
\begin{align}
\mathcal{P}_{\mathrm{c}}\left(\beta,\mathbf{x}_{\text{\ensuremath{\mathrm{tx}}}},\mathbf{x}_{\mathrm{rx}}\right) & =\underset{\triangleq\ \mathcal{P_{\mathrm{noint}}}}{\underbrace{\exp\bigl(-\beta^{\prime}\gamma_{\circ}\bigr)}}\times\underset{\triangleq\ \mathcal{P_{\circ}}\left(\Phi_{\mathrm{x}}\right)}{\underbrace{\mathbb{E}_{I_{\mathrm{x}}}\left\{ \exp\bigl(-\beta^{\prime}I\left(\Phi_{\mathrm{x}}\right)\bigr)\right\} }}\nonumber \\
 & \times\underset{\triangleq\ \mathcal{P_{\circ}}\left(\Phi_{\mathrm{y}}\right)}{\underbrace{\mathbb{E}_{I_{\mathrm{y}}}\left\{ \exp\bigl(-\beta^{\prime}I\left(\Phi_{\mathrm{y}}\right)\bigr)\right\} }},\label{eq: Success Probability: Pc =00003D Po * Px * Py}
\end{align}
where $\beta^{\prime}=\beta/\ell_{\mathrm{pl}}\left(\mathbf{x}_{\text{\ensuremath{\mathrm{tx}}}},\mathbf{x}_{\mathrm{rx}}\right)$,
$\mathcal{P_{\mathrm{noint}}}$ is the success probability in the
absence of interference, $\mathcal{P_{\circ}}\left(\Phi_{\mathrm{x}}\right)$
and $\mathcal{P_{\circ}}\left(\Phi_{\mathrm{y}}\right)$ are respectively
the degradation of the success probability due to independent aggregate
interference from the horizontal and vertical roads, where these two
factors can be evaluated by \cite{Moe_GC16}
\begin{align}
 & \mathcal{P_{\circ}}\left(\Phi_{\mathrm{q}}\right)=\exp\left(\!-\!\int_{\mathcal{B}}\frac{p_{\mathrm{I}}\lambda}{1+1/\left(\beta^{\prime}\ell_{\mathrm{pl}}\left(\mathbf{x},\mathbf{x}_{\mathrm{rx}}\right)\right)}\mathrm{d\bigl(\mathbf{x^{\mathrm{T}}\mathbf{e}_{\mathrm{q}}}\bigr)}\right),\,\mathrm{q}\in\{\mathrm{x},\mathrm{y}\}.\label{eq:  P(PPP) =00003D Theorem}
\end{align}

\subsection{\label{Sec3.3: Granular Reliability}Fine-grained Reliability}

The success probability $\mathcal{P}_{\mathrm{c}}\left(\beta,\mathbf{x}_{\text{\ensuremath{\mathrm{tx}}}},\mathbf{x}_{\mathrm{rx}}\right)$
provides a high-level performance assessment averaged over all possible
vehicular traffic realizations and channel. Unpacking this average
reliability and looking at it at the fine-grained or meta-level demands
that we study the success probability of the individual links. In
other words, we need to explore the \textit{meta distribution }of
the $\mathsf{SINR}$ \cite{Haenggi2016_MetaDistribution}; defined,
using the Palm probability $\mathbb{P}^{o}\left(\cdot\right)$, as
\begin{align}
F_{\mathrm{r}}\left(\beta,p\right) & \triangleq\mathbb{P}^{o}\Bigl(\thinspace\Pr\left(\mathsf{SINR}\geq\beta\thinspace|\thinspace\Phi_{\mathrm{x}}\cup\Phi_{\mathrm{y}}\right)\geq p\thinspace\Bigr)\label{eq: Meta Distribution (Success Probability)}
\end{align}
Introducing $p_{\mathrm{c}}\left(\beta\right)\triangleq\Pr\left(\mathsf{SINR}\geq\beta\thinspace|\thinspace\Phi_{\mathrm{x}}\cup\Phi_{\mathrm{y}}\right)$
and $p_{\mathrm{out}}\left(\beta\right)=1-p_{\mathrm{c}}\left(\beta\right)$
as the conditional success probability and conditional outage probability,
respectively, given the point process $\Phi_{\mathrm{x}}\cup\Phi_{\mathrm{y}}$,
and where $p\!\in\!\left[0,1\right]$ is a conditional success reliability
constraint. Thus, the meta distribution $F_{\mathrm{r}}\left(\beta,p\right)$
is the fraction of vehicular traffic realizations that achieve reliability,
where reliability is prescribed by the target value assigned to $p$,
and $\beta$ is the threshold for $\mathsf{SINR}$.

The average reliability of success can be obtained from $F_{\mathrm{r}}\left(\beta,p\right)$
as follows: using the Palm expectation $\mathbb{E}^{o}\left(\cdot\right)$,
we can obtain the moments of $p_{\mathrm{c}}\left(\beta\right)$,
the first being the average reliability of success, i.e., $\mathbb{E}^{o}\left(p_{\mathrm{c}}\left(\beta\right)\right)=\mathcal{P}_{\mathrm{c}}\left(\beta,\mathbf{x}_{\text{\ensuremath{\mathrm{tx}}}},\mathbf{x}_{\mathrm{rx}}\right)$.
Rather than obtaining the exact meta distribution through the Gil-Pelaez
theorem \cite{GilPelaez1951}, the moments could be used to approximate
$F_{\mathrm{r}}\left(\beta,p\right)$, where the beta distribution
(which only requires the first and second moments) is reported to
yield high accuracy \cite{Haenggi2016_MetaDistribution,Haenggi2017_MSalihi,Haenggi2017_MDWang}.

\section{\label{Sec4: Exact Reliability for Different Channels}Explicit Reliability
for Different Channels}

In this section, we are interested to tailor the average reliability
metrics for communications specified by the related vehicular propagation
models for suburban and urban intersections. 

\subsection{\label{Sec4.1: Reliability over Suburban Corners}Reliability for
Suburban Intersections}

The channel model of (\ref{eq: Channel - Suburban}) can be applied
to the success probability in (\ref{eq: Success Probability: Pc =00003D Po * Px * Py}).
The impact of random interferers on roads-$x$ and $y$ are accordingly
provided by the following propositions.

\newtheorem{prop1}{Proposition}
\begin{prop1}\label{MATH-Proposition: Suburban; Road-X}The function $\mathcal{P_{\circ}}\left(\Phi_{\mathrm{x}}\right)=\exp\left(-p_{\mathrm{I}}\lambda_{\mathrm{x}}\zeta_{\mathrm{s}}\mathcal{X}\left(R_{\mathrm{x}}\right)\right)$,
where $\mathcal{X}\left(R_{\mathrm{x}}\right)$ due to bounded PPP
interferers on road-$x$ tailored for suburban intersection evaluates
to 
\begin{align}
\mathcal{X}\left(R_{\mathrm{x}}\right) & =g_{\circ}\Bigl(\alpha,\bigl(R_{\mathrm{x}}+\left\Vert \mathbf{x}_{\mathrm{rx}}\right\Vert \bigr)/\zeta_{\mathrm{s}}\Bigr)\nonumber \\
 & +g_{\circ}\Bigl(\alpha,\bigl(R_{\mathrm{x}}-\left\Vert \mathbf{x}_{\mathrm{rx}}\right\Vert \bigr)/\zeta_{\mathrm{s}}\Bigr)\mathbf{1}_{\left\Vert \mathbf{x}_{\mathrm{rx}}\right\Vert \leq R_{\mathrm{x}}}\nonumber \\
 & -g_{\circ}\Bigl(\alpha,-\bigl(R_{\mathrm{x}}-\left\Vert \mathbf{x}_{\mathrm{rx}}\right\Vert \bigr)/\zeta_{\mathrm{s}}\Bigr)\mathbf{1}_{\left\Vert \mathbf{x}_{\mathrm{rx}}\right\Vert >R_{\mathrm{x}}},\label{eq: X(Rx) <--> Suburban, Road-X}
\end{align}
where $\zeta_{\mathrm{s}}=\beta^{1/\alpha}\left\Vert \mathbf{x}_{\mathrm{rx}}-\mathbf{x}_{\mathrm{tx}}\right\Vert $,
the function $g_{\circ}\left(\alpha,\vartheta\right)$ is defined
in (\ref{eqApp: A.4 =00005BG-Function=00005D}), and in which $\ensuremath{\mathbf{1}_{\mathsf{Q}}=1}$
when the statement $\ensuremath{\mathsf{Q}}$ is true and 0 otherwise.

\end{prop1}
\begin{IEEEproof}
See Appendix \ref{AppA: Proposition1}.
\end{IEEEproof}
\newtheorem{prop2}{}
\begin{prop1}\label{MATH-Proposition: Suburban; Road-Y}The function $\mathcal{P_{\circ}}\left(\Phi_{\mathrm{y}}\right)=\exp\left(-p_{\mathrm{I}}\lambda_{\mathrm{y}}\zeta_{\mathrm{s}}\mathcal{Y}\left(R_{\mathrm{y}}\right)\right)$,
where $\mathcal{Y}\left(R_{\mathrm{y}}\right)$ due to bounded PPP
interferers on road-$y$ tailored for suburban intersection evaluates
to 
\begin{align}
 & \mathcal{Y}\left(R_{\mathrm{y}}\right)=h_{\circ}\left(\alpha,\left(\left\Vert \mathbf{x}_{\mathrm{rx}}\right\Vert /\zeta_{\mathrm{s}}\right)^{2},\left(R_{\mathrm{y}}/\zeta_{\mathrm{s}}\right)^{2}\right),\label{eq: Y(Ry) <--> Suburban, Road-Y}
\end{align}
where $\zeta_{\mathrm{s}}=\beta^{1/\alpha}\left\Vert \mathbf{x}_{\mathrm{rx}}-\mathbf{x}_{\mathrm{tx}}\right\Vert $,
and the function $h_{\circ}\left(\alpha,\delta,\vartheta\right)$
is defined in (\ref{eqApp: B.3 =00005BH-Function=00005D}).

\end{prop1}
\begin{IEEEproof}
See Appendix \ref{AppB: Proposition2}.
\end{IEEEproof}
\begin{rem}
\noindent Studies have shown that in suburban environments, $\alpha=2$
is appropriate for inter-vehicular Euclidean distances below 100\,m,
and $\alpha=4$ for distances beyond \cite{Tufvesson2011IEEEMag}.
As is detailed in Appendix \ref{AppA: Proposition1}-\ref{AppB: Proposition2},
both $g_{\circ}\left(\alpha,\vartheta\right)$ and $h_{\circ}\left(\alpha,\delta,\vartheta\right)$
have exact closed-form expressions for $\alpha=2$.
\end{rem}
\begin{rem}
\noindent Rather than a bounded region, we may evaluate the success
probability impacted, in the aggregate, by infinite H-PPP interferers.
Say, we denote the degradation along roads-$x$ and $y$ by $\mathcal{P_{\circ}^{\infty}}\left(\Phi_{\mathrm{x}}\right)=\exp\left(-p_{\mathrm{I}}\lambda_{\mathrm{x}}\zeta_{\mathrm{s}}\mathcal{X}_{\infty}\right)$
and $\mathcal{P_{\circ}^{\infty}}\left(\Phi_{\mathrm{y}}\right)=\exp\left(-p_{\mathrm{I}}\lambda_{\mathrm{y}}\zeta_{\mathrm{s}}\mathcal{Y}_{\infty}\right)$,
where $\mathcal{X}_{\infty}$ and $\mathcal{Y}_{\infty}$ are determined
by taking the limit of (\ref{eq: X(Rx) <--> Suburban, Road-X}) and
(\ref{eq: Y(Ry) <--> Suburban, Road-Y}) as $R_{\mathrm{x}}$ and
$R_{\mathrm{y}}$ tend to infinity
\begin{align}
 & \!\!\mathcal{X}_{\infty}=2\pi\csc\bigl(\pi/\alpha\bigr)/\alpha\label{eq: X(inf) <--> Suburban, Road-X}\\
 & \!\!\mathcal{Y}_{\infty}=\begin{cases}
\lim_{\vartheta\rightarrow\infty}h_{\circ}\left(\alpha,\left(\left\Vert \mathbf{x}_{\mathrm{rx}}\right\Vert /\zeta_{\mathrm{s}}\right)^{2},\vartheta\right) & \left\Vert \mathbf{x}_{\mathrm{rx}}\right\Vert \neq0\\
2\pi\csc\bigl(\pi/\alpha\bigr)/\alpha & \left\Vert \mathbf{x}_{\mathrm{rx}}\right\Vert =0\mathrm{.}
\end{cases}\label{eq: Y(inf) <--> Suburban, Road-Y}
\end{align}
For all possible values of path loss exponent $\alpha$, the above
results are in fact congruent to the analysis reported in \cite{Erik_GC15}.
\end{rem}

\subsection{\label{Sec4.2: Reliability over Blind Urban Intersections}Reliability
for Blind Urban Intersections}

For the urban intersection, we proceed by plugging the path loss model
of (\ref{eq: Channel - Urban}) into (\ref{eq: Success Probability: Pc =00003D Po * Px * Py}).
The components along roads-$x$ and $y$ are provided by the following
propositions.\newtheorem{prop3}{}
\begin{prop1}\label{MATH-Proposition: Urban; Road-X}The function $\mathcal{P_{\circ}}\left(\Phi_{\mathrm{x}}\right)=\exp\left(-p_{\mathrm{I}}\lambda_{\mathrm{x}}\zeta_{\mathrm{u}}\mathcal{X}\left(R_{\mathrm{x}}\right)\right)$,
where $\mathcal{X}\left(R_{\mathrm{x}}\right)$ due to bounded $\mathrm{PPP}$
interferers on road-$x$ evaluated for urban intersection is  similar
to (\ref{eq: X(Rx) <--> Suburban, Road-X}), except the variable $\zeta_{\mathrm{s}}$
is replaced by
\begin{align}
 & \zeta_{\mathrm{u}}=\left(A_{\circ}\beta_{\mathrm{u}}^{\prime}\right)^{1/\alpha}=\Bigl(A_{\circ}\beta/\ell_{\mathrm{pl}}^{\mathrm{u}}\left(\mathbf{x}_{\mathrm{tx}},\mathbf{x}_{\mathrm{rx}}\right)\Bigr)^{1/\alpha},\label{eq: X(Rx) <--> Urban, Road-X}
\end{align}
where the channel model $\ell_{\mathrm{pl}}^{\mathrm{u}}\left(\mathbf{x}_{\mathrm{tx}},\mathbf{x}_{\mathrm{rx}}\right)$
is characterized in (\ref{eq: Channel - Urban}).

\end{prop1}
\begin{IEEEproof}
Similar to Appendix \ref{AppA: Proposition1}, with the exception
that the contribution from the wanted signal should take into account
the path loss model for the urban intersection with NLOS, WLOS or
LOS; thus, $\zeta_{\mathrm{s}}$ is replaced by $\zeta_{\mathrm{u}}$
shown in (\ref{eq: X(Rx) <--> Urban, Road-X}).
\end{IEEEproof}
\newtheorem{prop4}{}
\begin{prop1}\label{MATH-Proposition: Urban; Road-Y}The function $\mathcal{P_{\circ}}\left(\Phi_{\mathrm{y}}\right)=\exp\left(-p_{\mathrm{I}}\lambda_{\mathrm{y}}\zeta_{\mathrm{u}}\mathcal{Y}\left(R_{\mathrm{y}}\right)\right)$,
where $\mathcal{Y}\left(R_{\mathrm{y}}\right)$ due to bounded $\mathrm{PPP}$
interferers on road-$y$ tailored for urban intersection evaluates
to 
\begin{align}
\mathcal{Y}\left(R_{\mathrm{y}}\right) & =2\biggl(g_{\circ}\Bigl(\alpha,\bigl(R_{\mathrm{y}}+\left\Vert \mathbf{x}_{\mathrm{rx}}\right\Vert \bigr)/\zeta_{\mathrm{u}}\Bigr)\mathbf{1}_{\left\Vert \mathbf{x}_{\mathrm{rx}}\right\Vert \leq\triangle}\label{eq: Y(Ry) <--> Urban, Road-Y}\\
 & +\biggl(g_{\circ}\Bigl(\alpha,\bigl(\triangle+\left\Vert \mathbf{x}_{\mathrm{rx}}\right\Vert \bigr)/\zeta_{\mathrm{u}}\Bigr)+\frac{1}{\kappa}\biggl(g_{\circ}\Bigl(\alpha,\kappa R_{\mathrm{y}}/\zeta_{\mathrm{u}}\Bigr)\nonumber \\
 & -g_{\circ}\Bigl(\alpha,\kappa\triangle/\zeta_{\mathrm{u}}\Bigr)\biggr)\biggr)\mathbf{1}_{\left\Vert \mathbf{x}_{\mathrm{rx}}\right\Vert >\triangle}-g_{\circ}\Bigl(\alpha,\left\Vert \mathbf{x}_{\mathrm{rx}}\right\Vert /\zeta_{\mathrm{u}}\Bigr)\!\biggr),\nonumber 
\end{align}
where $\zeta_{\mathrm{u}}$ is given in (\ref{eq: X(Rx) <--> Urban, Road-X}),
$\kappa=\left(A_{\circ}/A_{\circ}^{\prime}\right)^{1/\alpha}\left\Vert \mathbf{x}_{\mathrm{rx}}\right\Vert $,
$\triangle$ is the break-point distance of the urban intersection
path loss, and the function $g_{\circ}\left(\alpha,\vartheta\right)$
is defined in (\ref{eqApp: A.4 =00005BG-Function=00005D}).

\end{prop1}
\begin{IEEEproof}
See Appendix \ref{AppC: Proposition4}.
\end{IEEEproof}
\begin{rem}
\noindent It is possible to show that degradation for an urban intersection
with infinite H-PPPs along roads-$x$ and $y$ is $\mathcal{P_{\circ}^{\infty}}\left(\Phi_{\mathrm{x}}\right)=\exp\left(-p_{\mathrm{I}}\lambda_{\mathrm{x}}\zeta_{\mathrm{u}}\mathcal{X}_{\infty}\right)$
and $\mathcal{P_{\circ}^{\infty}}\left(\Phi_{\mathrm{y}}\right)=\exp\left(-p_{\mathrm{I}}\lambda_{\mathrm{y}}\zeta_{\mathrm{u}}\mathcal{Y}_{\infty}\right)$,
where $\mathcal{X}_{\infty}$ is identical to (\ref{eq: X(inf) <--> Suburban, Road-X}),
and
\begin{align}
\mathcal{Y}_{\infty} & =2\biggl(\Bigl(\pi\csc\bigl(\pi/\alpha\bigr)/\alpha\Bigr)\mathbf{1}{}_{\left\Vert \mathbf{x}_{\mathrm{rx}}\right\Vert \leq\triangle}\label{eq: Y(inf) <--> Urban, Road-Y}\\
 & +\biggl(g_{\circ}\Bigl(\alpha,\bigl(\triangle+\left\Vert \mathbf{x}_{\mathrm{rx}}\right\Vert \bigr)/\zeta_{\mathrm{u}}\Bigr)+\frac{1}{\kappa}\biggl(\pi\csc\bigl(\pi/\alpha\bigr)/\alpha\nonumber \\
 & -g_{\circ}\Bigl(\alpha,\kappa\triangle/\zeta_{\mathrm{u}}\Bigr)\biggr)\biggr)\mathbf{1}{}_{\left\Vert \mathbf{x}_{\mathrm{rx}}\right\Vert >\triangle}-g_{\circ}\Bigl(\alpha,\left\Vert \mathbf{x}_{\mathrm{rx}}\right\Vert /\zeta_{\mathrm{u}}\Bigr)\!\biggr)\mathrm{.}\nonumber 
\end{align}
\end{rem}

\subsection{\label{Sec5: Analysis and Design}Network Analysis and Design}

To ensure a tolerable worst-case level of average performance, the
success probability must achieve a certain preassigned target value
$\mathcal{P}_{\mathrm{target}}\!\in\!\left(0,1\right)$, generally
close to 1, such that 
\begin{equation}
\mathcal{P_{\mathrm{noint}}}\mathcal{P_{\circ}}\left(\Phi_{\mathrm{x}}\right)\text{\ensuremath{\mathcal{P_{\circ}}\left(\Phi_{\mathrm{y}}\right)}}\geq\mathcal{P}_{\mathrm{target}}
\end{equation}
 over the intersection deployment region specified by $\mathcal{B}_{\mathrm{x}}\cup\mathcal{B}_{\mathrm{y}}$,
and for all V2V communications pair under consideration with positions
$\mathbf{x}_{\mathrm{tx}}$ and $\mathbf{x}_{\mathrm{rx}}$. As design
parameters, we consider the transmit probability $p_{\mathrm{I}}$
and its relation to road segments $R_{\mathrm{x}}$ and $R_{\mathrm{y}}$.
Solving for the transmit probability in this design criteria, we find
that $p_{\mathrm{I}}\leq p_{\mathrm{I}}^{\ast}\left(R_{\mathrm{x}},R_{\mathrm{y}}\right)$,
where the optimum probability is
\begin{align}
 & \!\!\!p_{\mathrm{I}}^{\ast}\left(R_{\mathrm{x}},R_{\mathrm{y}}\right)=\frac{\ln(\mathcal{P_{\mathrm{noint}}})-\ln(\mathcal{P}_{\mathrm{target}})}{\zeta\bigl(\lambda_{\mathrm{x}}\mathcal{X}\left(R_{\mathrm{x}}\right)+\lambda_{\mathrm{y}}\mathcal{Y}\left(R_{\mathrm{y}}\right)\bigr)},\label{eq: Pi(Rx,Ry) =00003D ...}
\end{align}
provided the natural condition $\mathcal{P}_{\mathrm{target}}\leq\mathcal{P}_{\mathrm{noint}}$
is met. Depending on the particular suburban or urban environment
under design, we assign in (\ref{eq: Pi(Rx,Ry) =00003D ...}) the
pertinent variables for $\zeta$ and the appropriate channel model
for $\ell_{\mathrm{pl}}\left(\mathbf{x}_{\mathrm{tx}},\mathbf{x}_{\mathrm{rx}}\right)$.

\begin{figure}[tbh]
\begin{centering}
\includegraphics[bb=0bp 0bp 526bp 425bp,width=0.75\columnwidth]{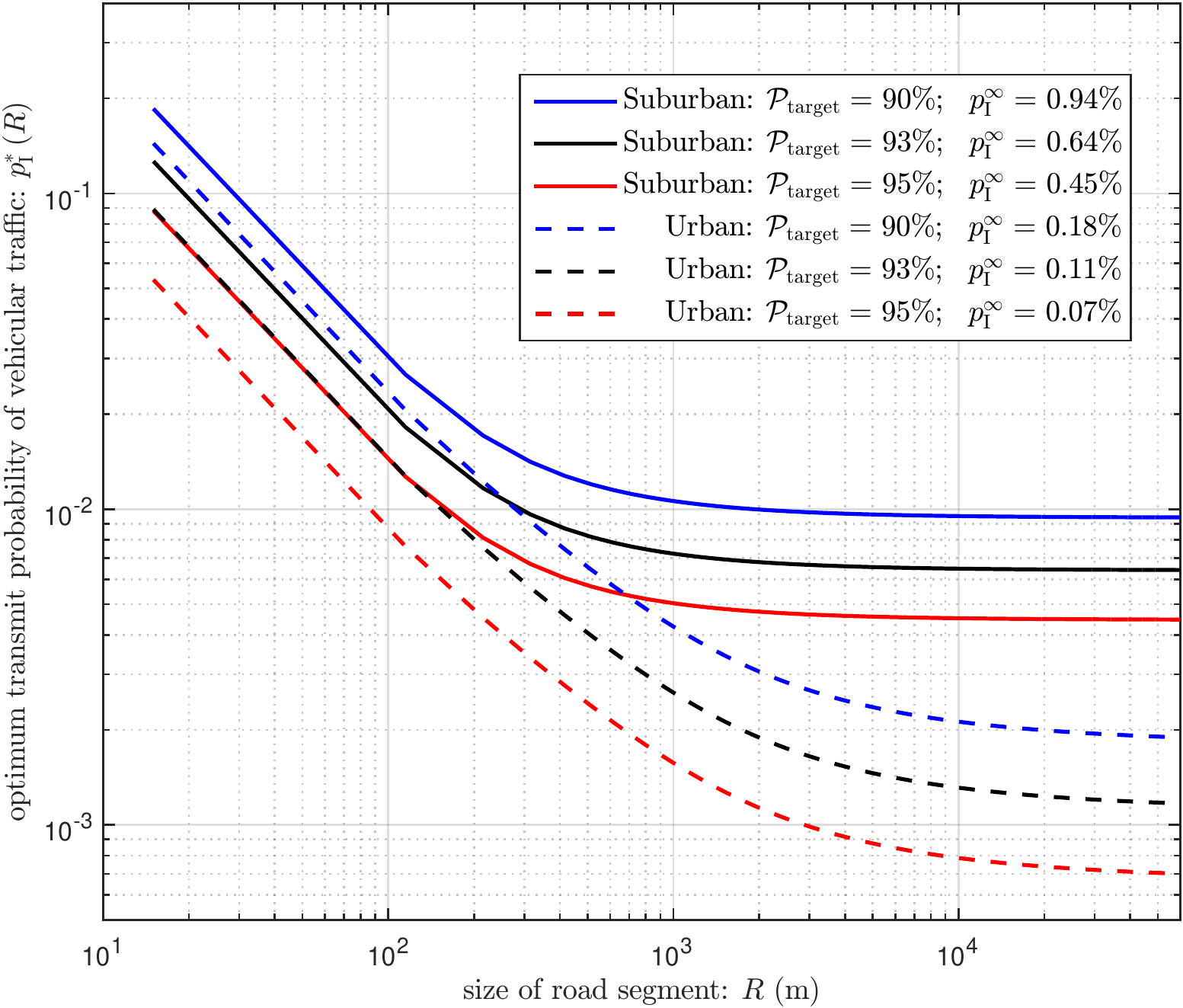}
\par\end{centering}
\centering{}\caption{Optimal transmit probability as a function of road segment $R\geq\triangle$,
over different values of target reliability $\mathcal{P}_{\mathrm{target}}$,
for suburban and urban channels. These curves are based on the worst-case
TX and RX positions for reliable V2V communications around the intersection,
i.e., $d_{\mathrm{max}}=100$\,m. The asymptotic transmit probability
$p_{\mathrm{I}}^{\infty}$ is associated to an infinite number of
interfering vehicles. \label{fig3: Network Design}}
\end{figure}

Evidently, it is possible to show that the minimum of $p_{\mathrm{I}}^{\ast}\left(R_{\mathrm{x}},R_{\mathrm{y}}\right)$
is associated with infinite H-PPPs evaluated with $\mathcal{X}_{\infty}$
and $\mathcal{Y}_{\infty}$, which we designate by $p_{\mathrm{I}}^{\infty}\triangleq\lim_{R_{\mathrm{x}},R_{\mathrm{y}}\rightarrow\infty}p_{\mathrm{I}}^{\ast}\left(R_{\mathrm{x}},R_{\mathrm{y}}\right)$.
Under the special case where the roads are of the same dimension,
i.e., $R_{\mathrm{x}}=R_{\mathrm{y}}=R$, the optimum transmit probability
simply becomes $p_{\mathrm{I}}^{\ast}\left(R\right)$, which we plot
in Fig.~\ref{fig3: Network Design} (with full simulation details
provided in the next section). We observe that $p_{\mathrm{I}}^{\ast}\left(R\right)$
is monotonically decreasing in $R$, since a larger region of possible
transmitters requires a reduction in the transmit probability in order
to achieve the target performance. It is interesting to note that
as $R$ becomes larger, the inverse proportionality reaches a plateau
characterized by a horizontal asymptote at $p_{\mathrm{I}}^{\infty}$.
 Also, as $\mathcal{P}_{\mathrm{target}}$ increases, the fraction
of active vehicles that can transmit simultaneous to the wanted TX/RX
link decreases. While keeping the same performance target, due to
the communications channel quality, more vehicles can be active in
the suburbs than in the city. Last, we should also be mindful that
the curves represent the optimum transmit probability, which means
that for a certain $R=R_{\circ}$, we could in fact choose any $p_{\mathrm{I}}$
below $p_{\mathrm{I}}^{\ast}\left(R_{\circ}\right)$ to meet the target.
Values above the curve do not meet the performance target.

\begin{table}[tbh]
\caption{Simulation Parameters\label{Tab: Simulation Parameters}}
\centering{}{\scriptsize{}}%
\begin{tabular}{ll}
\hline 
\multicolumn{2}{l}{\textbf{\scriptsize{}system parameters}}\tabularnewline
\hline 
\hline 
{\scriptsize{}target success probability} & {\scriptsize{}$\mathcal{P}_{\mathrm{target}}=0.9$}\tabularnewline
{\scriptsize{}transmit power} & {\scriptsize{}$P_{\circ}=20$\,dBmW}\tabularnewline
{\scriptsize{}AWGN floor} & {\scriptsize{}$N_{\circ}=-99$\,dBmW}\tabularnewline
{\scriptsize{}RX sensitivity} & {\scriptsize{}$\beta=8$\,dB (if $B=40$\,MHz; $r_{\mathrm{th}}\simeq115$\,Mbps)}\tabularnewline
{\scriptsize{}fixed Aloha transmit probability} & {\scriptsize{}$p_{\mathrm{I}}=0.02$\,(without network design)}\tabularnewline
\hline 
\hline 
\multicolumn{2}{l}{\textbf{\scriptsize{}channel propagation}}\tabularnewline
\hline 
\hline 
{\scriptsize{}operating frequency} & {\scriptsize{}$f_{\circ}=5.9$\,GHz}\tabularnewline
{\scriptsize{}reference distance} & {\scriptsize{}$d_{\circ}=10$\,m}\tabularnewline
{\scriptsize{}break-point distance} & {\scriptsize{}$\triangle=15$\,m}\tabularnewline
{\scriptsize{}path loss exponent} & {\scriptsize{}$\alpha=2$\,(suburban); $1.68$\,(urban)}\tabularnewline
{\scriptsize{}LOS/WLOS path loss coefficient} & {\scriptsize{}$A_{\circ}\!=\!-37.86+10\alpha$\,\,dBm}\tabularnewline
{\scriptsize{}NLOS path loss coefficient} & {\scriptsize{}$A_{\circ}^{\prime}\!=\!-38.32+\left(7\!+\!10\log_{10}\triangle\right)\!\alpha$\,\,dBm}\tabularnewline
\hline 
\hline 
\multicolumn{2}{l}{\textbf{\scriptsize{}vehicular traffic and geometry}}\tabularnewline
\hline 
\hline 
{\scriptsize{}traffic intensity} & {\scriptsize{}$\lambda=0.01$\,\#\,/\,m}\tabularnewline
{\scriptsize{}size of road segment} & {\scriptsize{}$R=200$\,m\,(practical); $10$\,km\,(stress-test)}\tabularnewline
{\scriptsize{}RX distance from junction point} & {\scriptsize{}$\left\Vert \mathbf{x}_{\mathrm{rx}}\right\Vert =50$\,m}\tabularnewline
{\scriptsize{}max. separation for reliable V2V com.} & {\scriptsize{}$d_{\mathrm{target}}=100$\,m\,\,($l_{1}$-norm distance)}\tabularnewline
{\scriptsize{}max. TX/RX Manhattan separation} & {\scriptsize{}$d_{\mathrm{max}}=140$\,m}\tabularnewline
\hline 
\hline 
\multicolumn{2}{l}{\textbf{\scriptsize{}Monte Carlo evaluation}}\tabularnewline
\hline 
\hline 
{\scriptsize{}reliability resolution} & {\scriptsize{}$m_{e}/d_{\mathrm{max}}=1$\,evaluation\,/\,m}\tabularnewline
{\scriptsize{}H-PPP realizations} & {\scriptsize{}$n_{\mathrm{PPP}}=10,000$}\tabularnewline
{\scriptsize{}fading iterations} & {\scriptsize{}$n_{\mathrm{f}}=5,000$}\tabularnewline
{\scriptsize{}histogram bins} & {\scriptsize{}$n_{\mathrm{b}}=150$}\tabularnewline
\hline 
\end{tabular}{\scriptsize \par}
\end{table}

\section{\label{Sec6: Simulations and Discussion}Simulations and Discussion}

\subsection{Simulation Setup}

Using the parameters shown in Table \ref{Tab: Simulation Parameters},
we evaluate the success probability under various conditions and scenarios.
In particular, we set the vehicular traffic on both roads to be the
same, i.e., $\lambda_{\mathrm{x}}\!=\!\lambda_{\mathrm{y}}\!=\!\lambda\!=\!0.01$\,\#/m.
For identical road segments $R_{\mathrm{x}}\!=\!R_{\mathrm{y}}\!=\!R$,
we consider $R\!\in\!\left\{ \mathcal{R}_{\mathrm{p}},\mathcal{R}_{\mathrm{s}}\right\} $
containing a subset of \textit{\emph{practical}}\emph{ }values for
real-world analysis and deployment: $\mathcal{R}_{\mathrm{p}}\!\in\!\left\{ 200,500\right\} $\,m;
and another subset utilized for \textit{\emph{stress-test}}\emph{
}analysis and fundamental limits: $\mathcal{R}_{\mathrm{s}}\!\in\!\left\{ 10,\infty\right\} $\,km.

Next, we assume a fixed RX on the horizontal, with $\mathbf{x}_{\mathrm{rx}}\!=\![-50,0]^{\mathrm{T}}$\,m;
and a TX that could take different positions, up to a Manhattan separation
of $d_{\mathrm{max}}\!=\!140$\,m away from the RX; i.e., curving
around the corner and going upward on the \textbf{$y$}-road. Due
to the nature of the urban channel model, we actually require, for
analysis and simulations, the spatial coordinates of the various TX
positions specified by $m_{\mathrm{e}}\!\in\!\mathbb{N}_{>0}$ equidistant
points, i.e.,
\begin{align}
 & \mathbf{x}_{\mathrm{tx}}\!\left(k\right)\!=\!\left(kd_{\mathrm{max}}/m_{\mathrm{e}}\!-\!\left\Vert \mathbf{x_{\mathrm{rx}}}\right\Vert \right)\left(\mathbf{1}_{k\in\mathcal{K}_{\mathrm{x}}}\mathbf{e}_{\mathrm{x}}\!+\!\mathbf{1}_{k\in\mathcal{K}_{\mathrm{y}}}\mathbf{e}_{\mathrm{y}}\right),\label{eq: Xtx(k) =00003D ...}
\end{align}
for $k\in\mathcal{K}_{\mathrm{x}}\!\cup\!\mathcal{K}_{\mathrm{y}}$,
and where $m_{x}\!\!=\!\!\bigl\lfloor m_{\mathrm{e}}\left\Vert \mathbf{x_{\mathrm{rx}}}\right\Vert /d_{\mathrm{max}}\bigr\rfloor$,
$\mathcal{K}_{\mathrm{x}}=\{k\in\mathbb{N}\thinspace\thinspace\bigl|\thinspace\thinspace1\leq k\leq m_{x};\;m_{x}\negmedspace>\negmedspace0\}$
and $\mathcal{K}_{\mathrm{y}}=\{k\in\mathbb{N}\thinspace\thinspace\bigl|\thinspace\thinspace\left(m_{x}+1\right)\leq k\leq m_{\mathrm{e}};\;m_{\mathrm{e}}\negmedspace>\negmedspace m_{x}\}$.

\subsection{Sensitivity of Average Reliability to TX/RX Separation}

Equipped with the above simulation setup, in this subsection, we evaluate
the sensitivity of the average success probability when the TX and
RX are in different locations. In particular, we design the network
by considering the guidelines specified by $p_{\mathrm{I}}^{\ast}\left(R\right)$,
while also taking into account the maximum Manhattan separation for
reliable V2V communications. That is, $d_{\mathrm{target}}\!=\!\left\Vert \mathbf{x}_{\mathrm{rx}}-\tilde{\mathbf{x}}_{\mathrm{tx}}\right\Vert _{1}\!=\!100$\,m,
where $\left\Vert \cdot\right\Vert _{1}$ denotes the $l_{1}$-norm,
and $\tilde{\mathbf{x}}_{\mathrm{tx}}$ is the position of the TX
at the target, which in this case corresponds to $\tilde{\mathbf{x}}_{\mathrm{tx}}\!=\!50\thinspace\mathbf{e}_{\mathrm{y}}$.
As a result, we inevitably anticipate that the average reliability
meets the target for different TX positions, i.e.,
\begin{align}
 & \mathcal{P}_{\mathrm{c}}\left(\beta,\mathbf{x}_{\mathrm{tx}},\mathbf{x}_{\mathrm{rx}}\right)\geq\mathcal{P}_{\mathrm{c}}\left(\beta,\tilde{\mathbf{x}}_{\mathrm{tx}},\mathbf{x}_{\mathrm{rx}}\right)=\mathcal{P}_{\mathrm{target}},\label{eq: Pc(x,x,b) >=00003D Design}
\end{align}
when the RX is fixed, and provided $\left\Vert \mathbf{x}_{\mathrm{rx}}-\mathbf{x}_{\mathrm{tx}}\right\Vert _{1}\leq\left\Vert \mathbf{x}_{\mathrm{rx}}-\tilde{\mathbf{x}}_{\mathrm{tx}}\right\Vert _{1}$.

\begin{figure*}[t]
\begin{centering}
\subfloat[sensitivity of outage probability over suburban corners\label{fig4a: Average Reliability - Suburban}]{\begin{centering}
\includegraphics[bb=0bp 0bp 526bp 425bp,width=0.5\columnwidth]{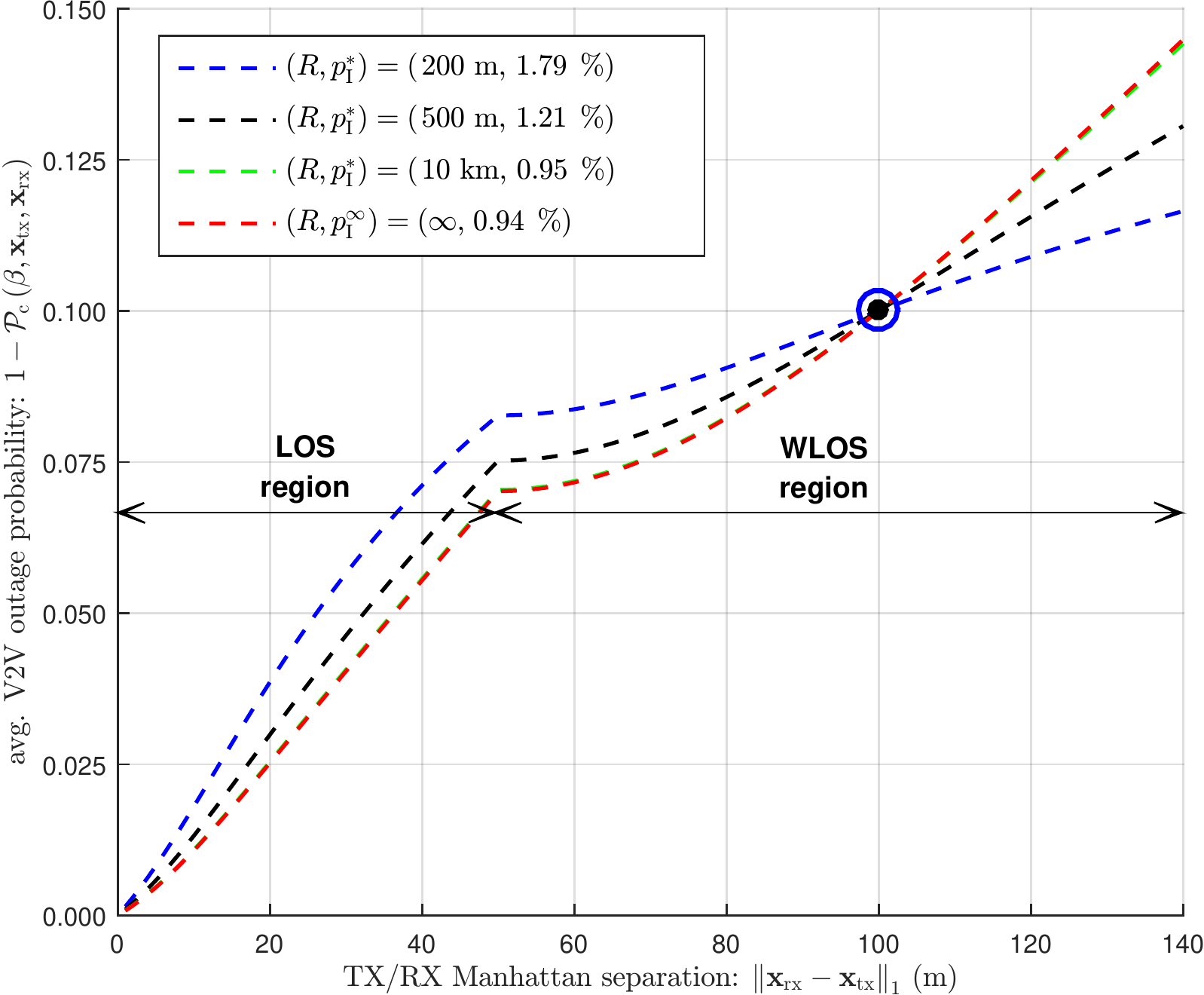}
\par\end{centering}
} \subfloat[sensitivity of outage probability over blind urban intersections\label{fig4b: Average Reliability - Urban}]{\begin{centering}
\includegraphics[bb=0bp 0bp 526bp 425bp,width=0.5\columnwidth]{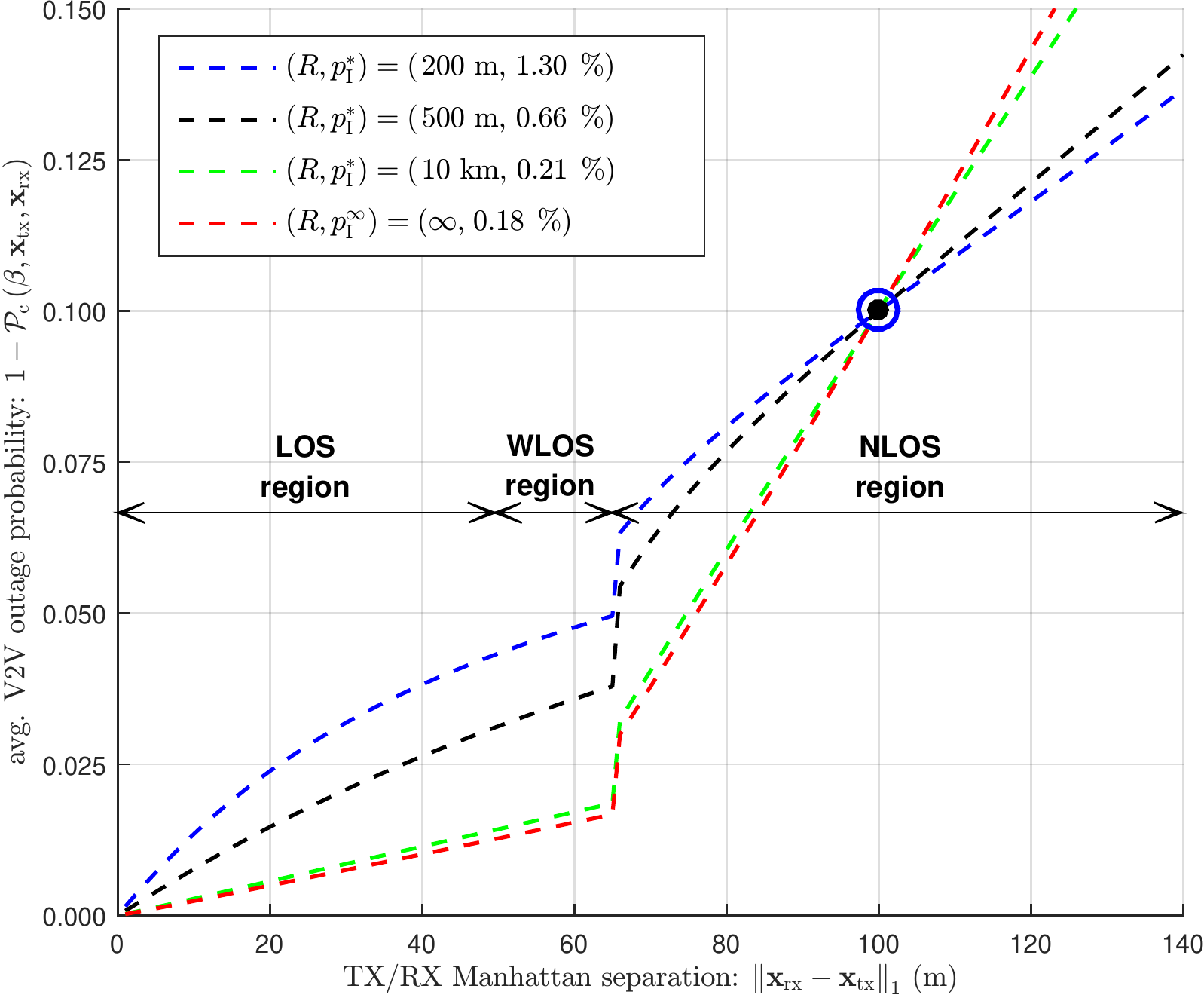}
\par\end{centering}
}
\par\end{centering}
\centering{}\caption{Average reliability as a function of TX/RX separation for different
channel environments and network design choices; whereby the optimum
transmit probability $p_{\mathrm{I}}^{\ast}\left(R\right)$ is designed
to meet a target reliability of $\mathcal{P}_{\mathrm{target}}=0.9$,
at the largest V2V separation of $d_{\mathrm{target}}=100$\,m for
reliable communications.\label{fig4: Average Reliability}}
\end{figure*}

For visualization purposes, we plot in Fig.~\ref{fig4: Average Reliability},
the average outage probability as a function of TX/RX Manhattan separation,
designed\footnote{For other design criteria, such as $d_{\mathrm{target}}\!\in\!\bigl\{20,40,60,80,120\bigr\}$\,m,
which equivalently corresponds to $\tilde{\mathbf{x}}_{\mathrm{tx}}\!\in\!\bigl\{-30\thinspace\mathbf{e}_{\mathrm{x}},-10\thinspace\mathbf{e}_{\mathrm{x}},10\thinspace\mathbf{e}_{\mathrm{y}},30\thinspace\mathbf{e}_{\mathrm{y}},70\thinspace\mathbf{e}_{\mathrm{y}}\bigr\}$,
sensitivity in an urban channel is available in \cite{Moe_GC16}.} for $d_{\mathrm{target}}\!\!=\!\!100$\,m, sweeping different road
segments $R$, and over suburban and urban channel environments. We
first note that when $\left\Vert \mathbf{x}_{\mathrm{rx}}-\mathbf{x}_{\mathrm{tx}}\right\Vert _{1}\!=\!d_{\mathrm{target}}$,
a reliability of $0.9$ (shown with a blue circle mark, for a corresponding
outage of $0.1$) is achieved. When $\left\Vert \mathbf{x}_{\mathrm{rx}}-\mathbf{x}_{\mathrm{tx}}\right\Vert _{1}\!<\!d_{\mathrm{target}}$,
the outage reduces, while for $\left\Vert \mathbf{x}_{\mathrm{rx}}-\mathbf{x}_{\mathrm{tx}}\right\Vert _{1}\!>\!d_{\mathrm{target}}$,
the outage increases.
 We note that the smallest interference region (i.e., $R\!=\!200$\,m)
corresponds to the largest transmit probability at target. Also, this
smallest interference region leads to the \textit{largest} outages
for $\left\Vert \mathbf{x}_{\mathrm{rx}}-\mathbf{x}_{\mathrm{tx}}\right\Vert _{1}\!<\!d_{\mathrm{target}}$,
though never surpassing $0.1$. This is due to the larger possibility
of active transmitters in close proximity to the typical user, i.e.,
the RX. On the other hand, the smallest interference region leads
to the \textit{smallest} outages for $\left\Vert \mathbf{x}_{\mathrm{rx}}-\mathbf{x}_{\mathrm{tx}}\right\Vert _{1}\!>\!d_{\mathrm{target}}$.
This is because the outage is dominated by the aggregate interference,
rather than the interferers close to the RX. Hence, the larger deployment
region, which has more interferers, has the largest outages. Ultimately,
it is important to emphasize that a small interference region allows
for a high density of active transmitters $\lambda p_{\mathrm{I}}^{\ast}\left(R\right)$,
while leading to relatively graceful degradation outside the interference
region. We also notice that when we consider a very large deployment
region, say $R\!=\!10$\,km, the curves nearly overlap the fundamental
case for infinite region, and so this value is a feasible choice for
stress-test analysis.

The outage sensitivity varies with a distinctive format as the TX
moves across different regions, which is consistent with the uniqueness
of the intersection path loss models. For the suburban case, there
are two regions: (i) LOS: $\left\Vert \mathbf{x}_{\mathrm{rx}}-\mathbf{x}_{\mathrm{tx}}\right\Vert _{1}\!\!\in\!\!\bigl(0,\left\Vert \mathbf{x}_{\mathrm{rx}}\right\Vert \bigr]$;
and (ii) WLOS: $\left\Vert \mathbf{x}_{\mathrm{rx}}-\mathbf{x}_{\mathrm{tx}}\right\Vert _{1}\!\!\in\!\!\bigl(\left\Vert \mathbf{x}_{\mathrm{rx}}\right\Vert ,d_{\mathrm{max}}\bigr]$.
On the other hand, for the urban case, the TX undergoes three regions:
(i) LOS: $\left\Vert \mathbf{x}_{\mathrm{rx}}-\mathbf{x}_{\mathrm{tx}}\right\Vert _{1}\!\!\in\!\!\bigl(0,\left\Vert \mathbf{x}_{\mathrm{rx}}\right\Vert \bigr]$;
(ii) WLOS: $\left\Vert \mathbf{x}_{\mathrm{rx}}-\mathbf{x}_{\mathrm{tx}}\right\Vert _{1}\!\!\in\!\!\bigl(\left\Vert \mathbf{x}_{\mathrm{rx}}\right\Vert ,\left\Vert \mathbf{x}_{\mathrm{rx}}\right\Vert +\triangle\bigr]$;
and (iii) NLOS: $\left\Vert \mathbf{x}_{\mathrm{rx}}-\mathbf{x}_{\mathrm{tx}}\right\Vert _{1}\!\!\in\!\!\bigl(\left\Vert \mathbf{x}_{\mathrm{rx}}\right\Vert +\triangle,d_{\mathrm{max}}\bigr]$.
As expected, the results reveal a significant deterioration of the
communications reliability as the TX transitions from LOS to WLOS,
and towards NLOS. Moreover, due to the non-continuous nature of the
urban channel model in (\ref{eq: Channel - Urban}), the outage curves
show an abrupt transition when $\left\Vert \mathbf{x}_{\mathrm{rx}}-\mathbf{x}_{\mathrm{tx}}\right\Vert _{1}=\left\Vert \mathbf{x}_{\mathrm{rx}}\right\Vert +\triangle$,
i.e. at the passing from the WLOS to the NLOS region, which in this
case corresponds to a separation of $65$\,m.

Overall, our analysis of the average reliability indicates that when
a system is designed for a certain maximum communication range (e.g.,
$d_{\mathrm{target}}\!=\!100$\,m), it is recommended to set the
deployment region $R$ as low as possible (in this case, $R\!=\!200$\,m
is recommended), as this leads to the highest density of active transmitters
and a graceful performance degradation outside the interference region
(i.e., when $R\!>\!200$\,m). Furthermore, for a particular target
performance, a better channel environment with less hindering obstacles,
such as the case with suburban corners, can tolerate more active interfering
nodes (i.e., a larger $p_{\mathrm{I}}^{\ast}$) than in built-up blind
urban intersections, where the transmission quality is severely degraded
over a relatively short communications distance.

\subsection{Sensitivity of Meta Distribution to Network Design}

In the previous subsection, we analyzed the reliability averaged over
point processes of the vehicular traffic and the randomness of the
channel. Examining this average reliability metric is insightful,
yet it is incomplete. For instance, say we consider $n_{\mathrm{PPP}}\!\!=\!10,000$
traffic realizations, we are then intrigued to know: \textit{how many
of these realizations actually meet the preset requirement for target
reliability $\mathcal{P}_{\mathrm{target}}$?} The meta distribution
defined in (\ref{eq: Meta Distribution (Success Probability)}) quantifies
the fraction of traffic realizations that achieve a reliability constraint.
In other words, the answer to the above question is: $\left\lfloor n_{\mathrm{PPP}}\thinspace F_{\mathrm{r}}\left(\beta,\mathcal{P}_{\mathrm{target}}\right)\right\rfloor $.
\begin{figure*}[t]
\begin{centering}
\subfloat[meta distribution without network design: $p_{\mathrm{I}}=0.02$\label{fig5a: MD - NoDesign}]{\begin{centering}
\includegraphics[bb=0bp 0bp 526bp 425bp,width=0.5\columnwidth]{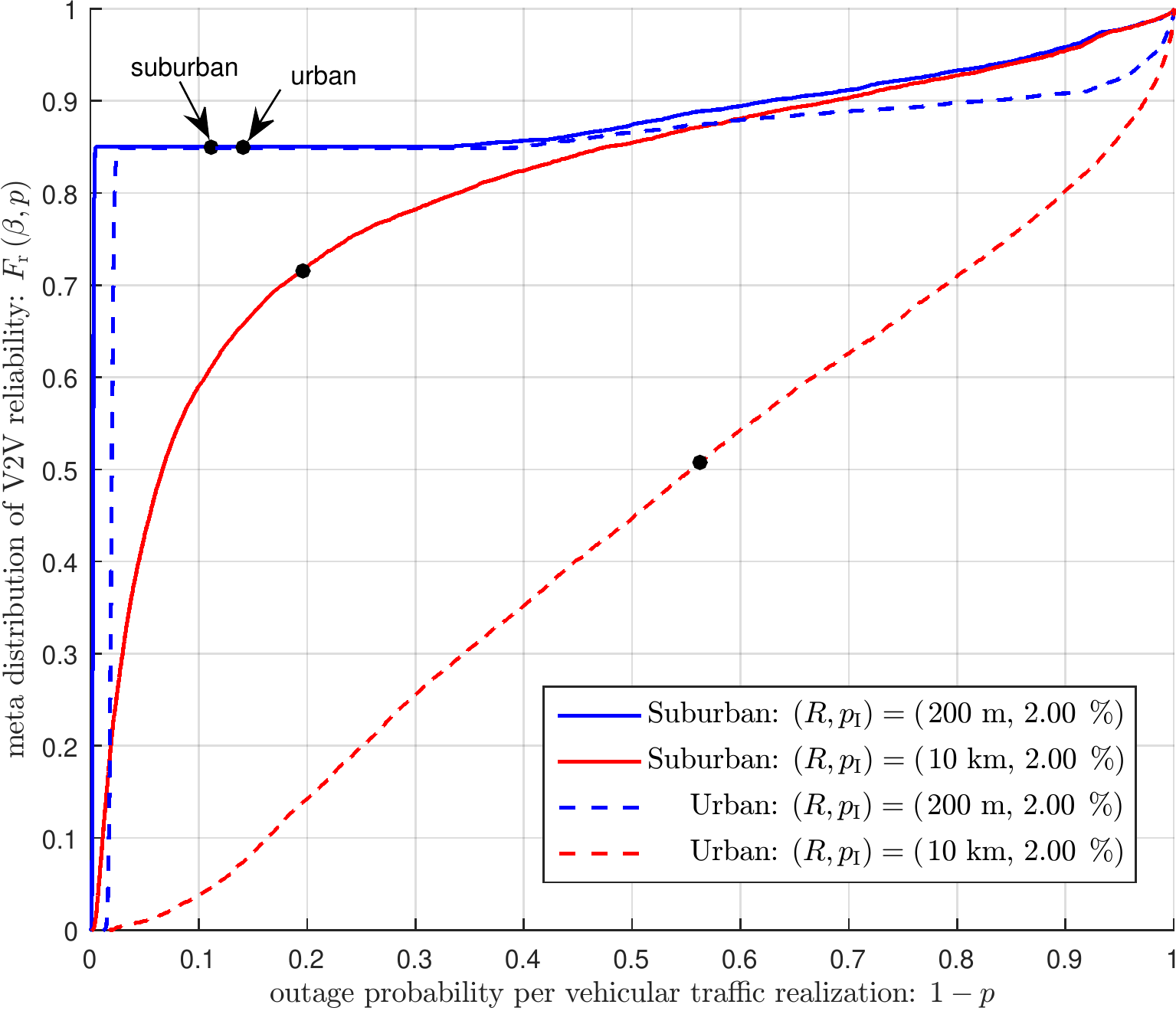}
\par\end{centering}
} \subfloat[meta distribution with network design: $p_{\mathrm{I}}^{\ast}\left(R\right)$\label{fig5b: MD - Design}]{\begin{centering}
\includegraphics[bb=0bp 0bp 526bp 425bp,width=0.5\columnwidth]{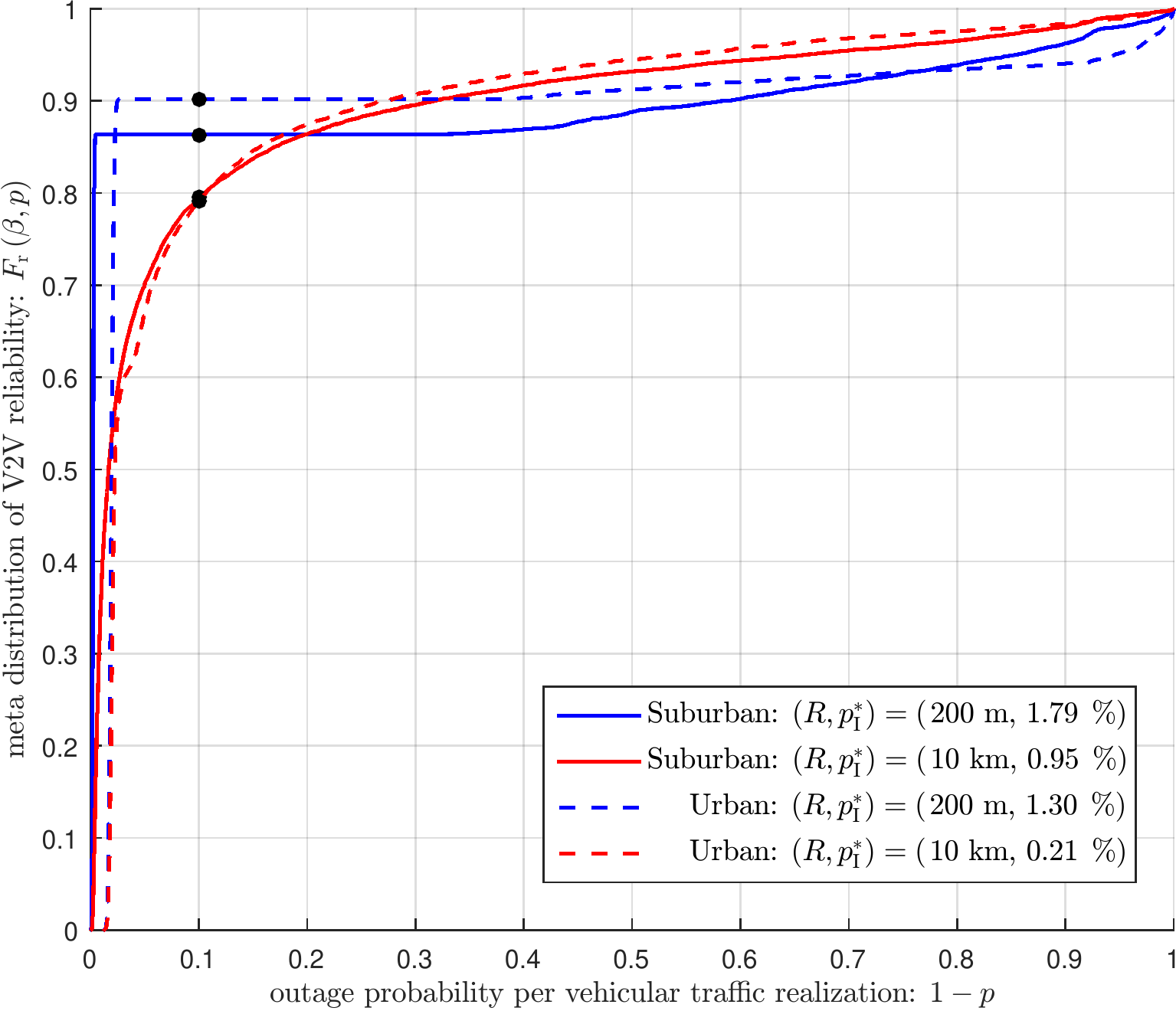}
\par\end{centering}
}
\par\end{centering}
\centering{}\caption{Meta distribution of reliability as a function of outage probability
conditioned on vehicular traffic realization under two network scenarios:
(a) without design, and (b) with design. For each channel environment,
the fraction of vehicular traffic is shown for $R=200$\,m\,(practical)
and $R=10$\,km\,(stress-test). TX and RX are on orthogonal roads,
$50$\,m away from the junction. The dots on the distribution curves
represent the Palm expectation.  \label{fig5: MD - NoDesign  -or-  Design}}
\end{figure*}

To understand the sensitivity to network design, we consider the meta
distribution for two cases: (i) without the application of network
design, i.e., transmit probability is fixed at all times to $p_{\mathrm{I}}\!=\!0.02$;
and (ii) with network design guidelines applied through $p_{\mathrm{I}}^{\ast}\left(R\right)$.
The setup for the simulation is similar as before, and where the TX
and RX are both fixed $50$\,m away from the junction. Of course,
the random vehicular traffic on the intersection roads vary for each
of the $n_{\mathrm{PPP}}\!\!=\!10,000$ realizations. Meanwhile, for
each realization, we determine the related outage value, estimated
via Monte Carlo techniques based on $n_{\mathrm{f}}\!=\!5,000$ fading
iterations. Once this process is completed for the $10,000$ realizations,
we then assemble the outage data and build an $n_{\mathrm{b}}\!=\!150$
bin histogram for the probability mass function, from which a CDF
plot of the meta distribution is obtained. This is shown in Fig.~\ref{fig5: MD - NoDesign  -or-  Design}
for different design options, channels, and road segments. From analysis,
we know that $\mathbb{E}^{o}\left(p_{\mathrm{out}}\left(\beta\right)\right)\!=\!1\!-\!\mathbb{E}^{o}\left(p_{\mathrm{c}}\left(\beta\right)\right)\!=\!1\!-\!\mathcal{P}_{\mathrm{c}}\left(\beta,\mathbf{x}_{\text{\ensuremath{\mathrm{tx}}}},\mathbf{x}_{\mathrm{rx}}\right)$,
shown as dots in Fig.~\ref{fig5: MD - NoDesign  -or-  Design}. 

Comparing the subfigures in Fig.~\ref{fig5: MD - NoDesign  -or-  Design},
we observe that when network design is not applied, the average outage
$1-\mathcal{P}_{\mathrm{c}}\left(\beta,\mathbf{x}_{\text{\ensuremath{\mathrm{tx}}}},\mathbf{x}_{\mathrm{rx}}\right)$
varies from 0.56 (urban, $R=10\,\mathrm{km}$) to 0.11 (suburban,
$R=200\,\mathrm{m}$). The corresponding meta distribution $F_{\mathrm{r}}\left(\beta,\mathcal{P}_{\mathrm{c}}\left(\beta,\mathbf{x}_{\text{\ensuremath{\mathrm{tx}}}},\mathbf{x}_{\mathrm{rx}}\right)\right)$
varies between 0.51 to 0.85, indicating that for the urban intersection
with $R=10\,\mathrm{km}$ about half the PPP realizations achieve
the target reliability $\mathcal{P}_{\mathrm{c}}\left(\beta,\mathbf{x}_{\text{\ensuremath{\mathrm{tx}}}},\mathbf{x}_{\mathrm{rx}}\right)$,
while this number increases to nearly 90\% for the suburban intersection
at $R=200\,\mathrm{m}$. In other words, for small value of $R$,
the meta-distribution exhibits a bimodal behavior, characterized by
a large number of ``good'' PPPs and a small number of ``bad''
PPPs, meaning the traffic realizations lead to either very reliable
or extremely unreliable communication conditions. When network design
is applied (see Fig.~\ref{fig5b: MD - Design}), $\mathbb{E}^{o}\left(p_{\mathrm{out}}\left(\beta\right)\right)=1-\mathcal{P}_{\mathrm{c}}\left(\beta,\mathbf{x}_{\text{\ensuremath{\mathrm{tx}}}},\mathbf{x}_{\mathrm{rx}}\right)=1-\mathcal{P}_{\mathrm{target}}=0.1$,
while $F_{\mathrm{r}}\left(\beta,\mathcal{P}_{\mathrm{target}}\right)=F_{\mathrm{r}}\left(\beta,0.9\right)$
ranges from 0.8 to 0.9, meaning that a greater fraction of traffic
realizations achieve the target reliability when compared to the case
of no design. Note that when $R\!=\!200$\,m in Fig.~\ref{fig5b: MD - Design},
the meta distribution for the urban channel is slightly higher than
for the suburban channel since, due to design, the suburban case allows
a higher value of transmit probability, i.e., more simultaneously
active vehicles are tolerated. Moreover, for $R\!=\!200$\,m, the
meta distribution is again bimodal, while for $R=10\,\mathrm{km}$
it is not.  Overall, these results indicate that the average performance
alone is not an adequate metric to assess communication reliability,
but with different reasons for large and small transmission ranges. 

\begin{figure*}[tbh]
\begin{centering}
\subfloat[fine-grained reliability, suburban corners: $R=200$\,m\label{fig6a: Granular Reliability - Suburban, R=00003D200m--}]{\begin{centering}
\includegraphics[bb=0bp 0bp 526bp 425bp,width=0.5\columnwidth]{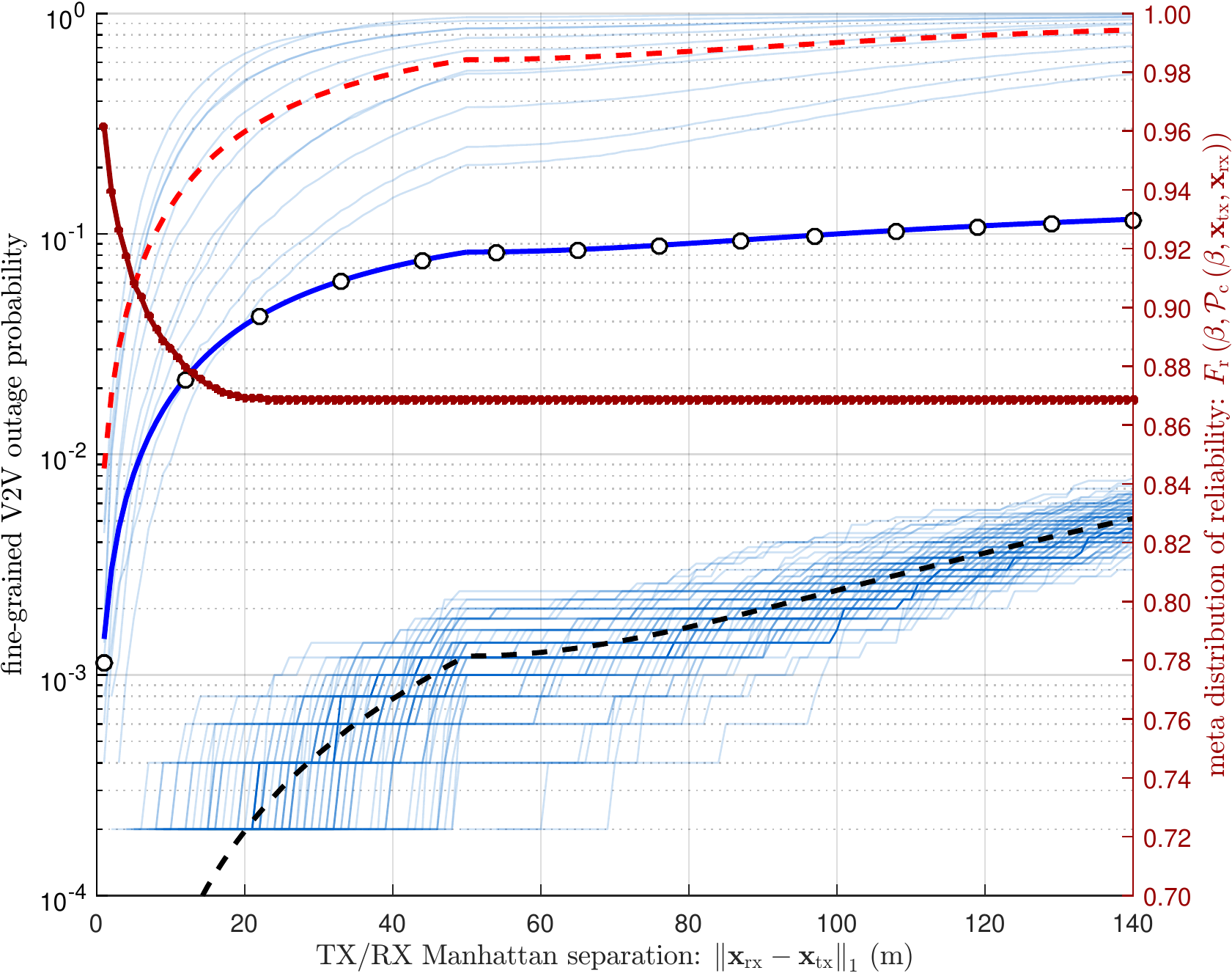}
\par\end{centering}
} \subfloat[fine-grained reliability, blind urban intersections: $R=200$\,m\label{fig6b: Granular Reliability - Urban, R=00003D200m--}]{\begin{centering}
\includegraphics[bb=0bp 0bp 526bp 425bp,width=0.5\columnwidth]{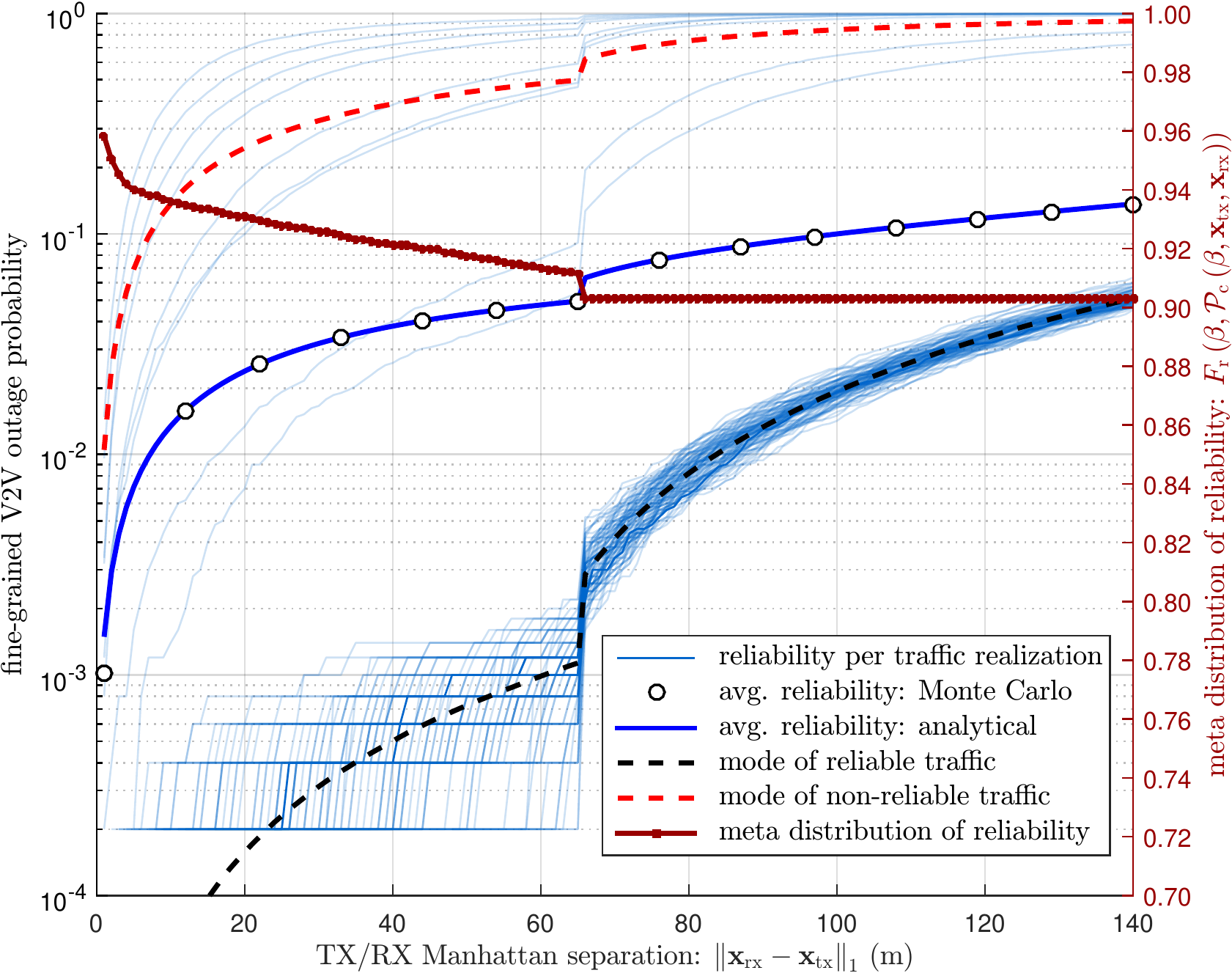}
\par\end{centering}
}
\par\end{centering}
\begin{centering}
\subfloat[fine-grained reliability, suburban corners: $R=10$\,km\label{fig6c: Granular Reliability - Suburban, R=00003D10km--}]{\begin{centering}
\includegraphics[bb=0bp 0bp 526bp 425bp,width=0.5\columnwidth]{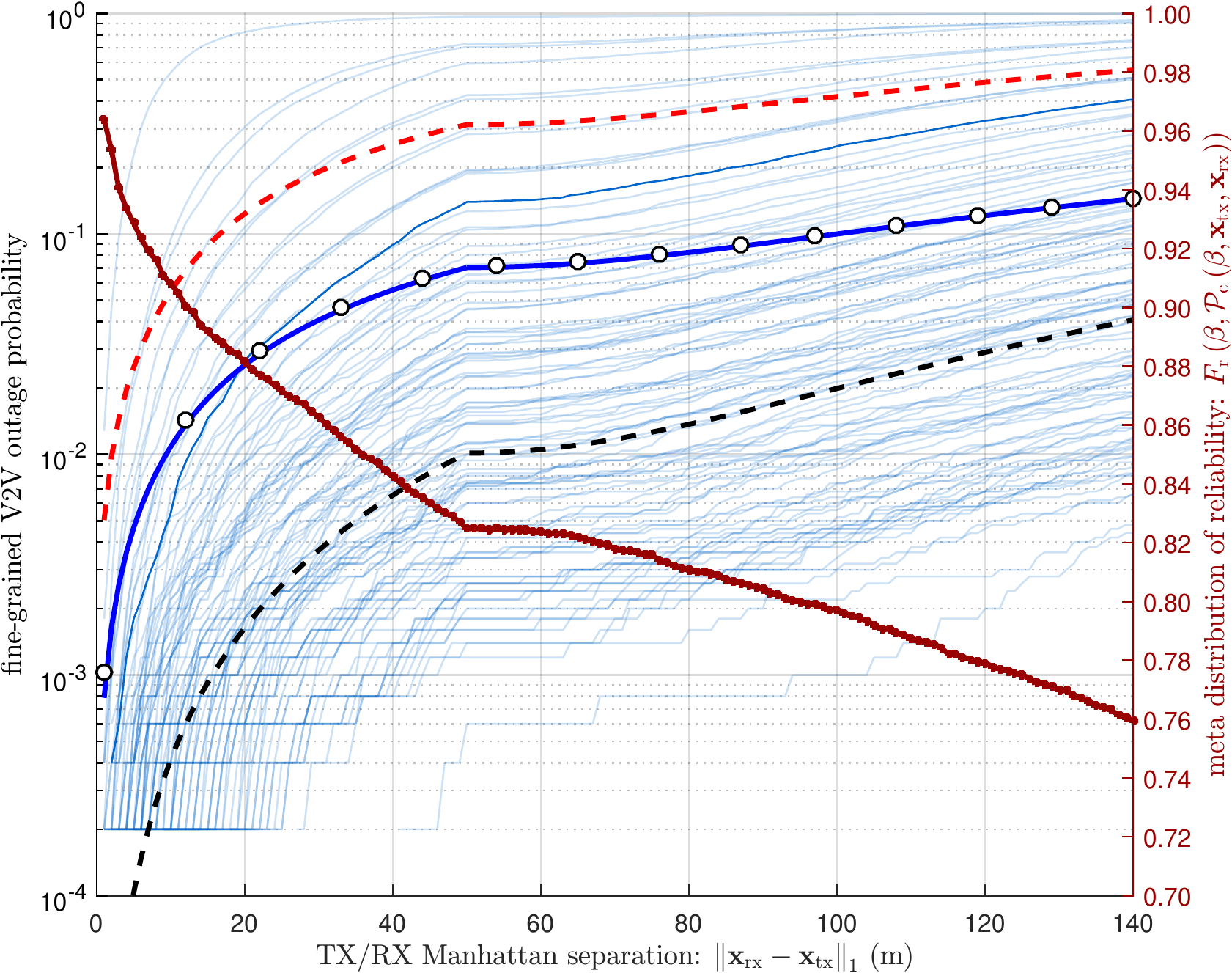}
\par\end{centering}
} \subfloat[fine-grained reliability, blind urban intersections: $R=10$\,km\label{fig6d: Granular Reliability - Urban, R=00003D10km--}]{\begin{centering}
\includegraphics[bb=0bp 0bp 526bp 425bp,width=0.5\columnwidth]{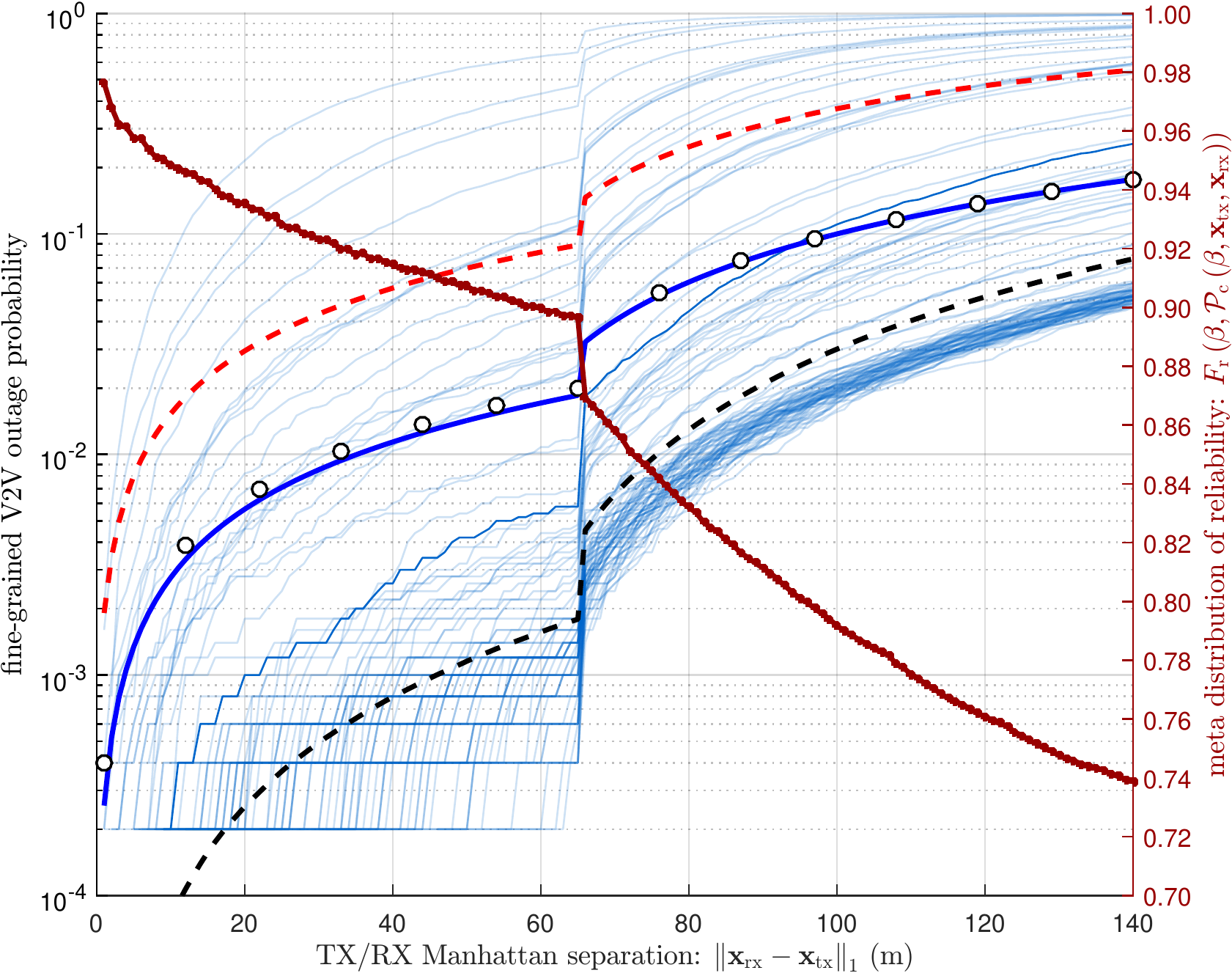}
\par\end{centering}
}
\par\end{centering}
\centering{}\caption{Fine-grained reliability under network design as a function of TX/RX
separation for different channel environments and road segments. The
analysis is based on $n_{\mathrm{ppp}}=10,000$, however for illustration
purposes, we only show the first $100$ fine-grained outage curves
evaluated using Monte Carlo simulation, where each fine-line is associated
with a certain vehicular traffic realization. The average reliability
is then estimated and compared to the analytically derived expressions.
The mode for reliable and non-reliable traffic identified at target
reliability $\left(d_{\mathrm{target}},1-\mathcal{P}_{\mathrm{target}}\right)=\left(100\thinspace\mathrm{m},0.1\right)$
are plotted. The meta distribution of reliability is also shown for
different V2V separations.\label{fig6: Granular Reliability--}}
\end{figure*}

\subsection{Sensitivity of Fine-grained Reliability to TX/RX Separation}

In the previous subsection, we looked at the meta distribution of
reliability. Here, we also aim to obtain the meta distribution as
a function of $\mathbf{x}_{\text{\ensuremath{\mathrm{tx}}}}$ and
$\mathbf{x}_{\mathrm{rx}}$, obtained from the fine-grained reliability
plots, i.e., outage probability per traffic realization as a function
of TX/RX Manhattan separation. We consider the same setup explained
in the previous section (now designed for $d_{\mathrm{target}}=100$\,m
and $\mathcal{P}_{\mathrm{target}}=0.9$) to obtain an outage value
for each realization. However, here, the RX will remain fixed at $\mathbf{x}_{\mathrm{rx}}\!=\!-50\thinspace\mathbf{e}_{\mathrm{x}}$,
and the TX will have different positions with a reliability resolution
of $m_{e}/d_{\mathrm{max}}=1$\,evaluation\,/\,m. For every TX/RX
position pair, we obtain $10,000$ outage values. Therefore, we have
to do this process over many times to get a fine-grained reliability,
i.e., $m_{e}\times n_{\mathrm{ppp}}=140\times10,000=1.4$\,Million
outage values that must be estimated, and where each of these values
is averaged over $n_{\mathrm{f}}=5,000$ fading iterations. 

Similar to Fig.~\ref{fig4: Average Reliability}, in Fig.~\ref{fig6: Granular Reliability--},
we plot the average outage $1-\mathcal{P}_{\mathrm{c}}\left(\beta,\mathbf{x}_{\text{\ensuremath{\mathrm{tx}}}},\mathbf{x}_{\mathrm{rx}}\right)$
(with an axis on the left), and the meta distribution $F_{\mathrm{r}}\left(\beta,\mathcal{P}_{\mathrm{c}}\left(\beta,\mathbf{x}_{\text{\ensuremath{\mathrm{tx}}}},\mathbf{x}_{\mathrm{rx}}\right)\right)$
(with an axis on the right) both as a function of $\left\Vert \mathbf{x}_{\mathrm{rx}}-\mathbf{x}_{\mathrm{tx}}\right\Vert _{1}$.
With increasing $\left\Vert \mathbf{x}_{\mathrm{rx}}-\mathbf{x}_{\mathrm{tx}}\right\Vert _{1}$,
$1-\mathcal{P}_{\mathrm{c}}\left(\beta,\mathbf{x}_{\text{\ensuremath{\mathrm{tx}}}},\mathbf{x}_{\mathrm{rx}}\right)$
increases, while $F_{\mathrm{r}}\left(\beta,\mathcal{P}_{\mathrm{c}}\left(\beta,\mathbf{x}_{\text{\ensuremath{\mathrm{tx}}}},\mathbf{x}_{\mathrm{rx}}\right)\right)$
decreases. Both of these phenomena are a natural V2V communications
response, as the TX vehicle transitioning from the LOS region to the
WLOS region, and finally onto the NLOS region around the blind-intersection.
In other words, the overall trend for average reliability and fine-grained
reliability indicate a deterioration of the communications quality.
Also, in Figs.~\ref{fig6a: Granular Reliability - Suburban, R=00003D200m--}\textendash \ref{fig6b: Granular Reliability - Urban, R=00003D200m--},
after a certain TX/RX separation, the meta distribution reaches a
steady-state, which is a clear indication of the bimodal behavior
for small values of road segments.

Although averaged over $10,000$ realizations, for visualization purposes,
we only show the fine-grained reliability for 100 randomly drawn traffic
realizations. The bimodal behavior for $R=200\,\mathrm{m}$ is again
clearly visible: we have either very reliable traffic realizations
or (by design, fewer) non-reliable realizations, but nowhere in close
proximity to the average measure. For the more realistic range of
$R=200\,\mathrm{m}$, outage is mainly determined by the presence
or absence of interferers, while for $R=10\,\mathrm{km}$, outage
is mainly determined by the distance to the closest interferers. Grouping
the 10,000 realizations based on whether or not they meet the target,
we can compute conditional outage probabilities, shown as dashed lines.
For $R=10\,\mathrm{km}$, these lines are a measure of the spread
of the meta distribution around the mean, while for $R=200\,\mathrm{m}$,
the dashed lines are an indication of the separation between the two
modes of the meta distribution.

Overall, we find that the generally accepted notion that reliability
for a particular traffic realization will be in relative proximity
to the average is utterly misleading. In fact, in this case, averages
are essentially an oversimplified distortion of reality. They provide
ballpark values, but these values may not necessarily exist for a
particular vehicular traffic scenario as shown here.

\section{\label{Sec7: Conclusion}Conclusion}

V2V communication is critical for future intelligent transportation
systems. A key performance metric is the probability of successful
packet delivery in the presence of interference. In this paper, we
analytically characterized the success probability for suburban and
urban intersections based on specialized path loss models. It turns
out that these path loss models are amenable for mathematical analysis
and lead to exact closed-form expressions for different path loss
exponents and finite interference regions. The derived expressions
can aid in the communication system design task, complementing time-consuming
simulations and experiments. In particular, we found that from a system
perspective, it is beneficial to limit interference to a small spatial
region, while allowing more simultaneous transmitters. We also explained
the notion of network design in order to ascertain a target reliability.
This was coupled with the meta distribution, where it was shown that
a small road segment can lead to a higher fraction of traffic realizations
that achieve the target reliability, as compared to infinitely long
roads. Finally, fine-grained reliability per traffic realization reveals
a bimodal distribution outcome for practical real-world deployment
over short road segments, thus indicating that traditional metrics
based on averages are important, but not sufficient, for ultra-reliable
applications such as V2V communications.

\appendices{}

\numberwithin{equation}{section}

\section{\label{AppA: Proposition1}Proof of Proposition \ref{MATH-Proposition: Suburban; Road-X}:
$\mathcal{P_{\circ}}\left(\Phi_{\mathrm{x}}\right)$ for Suburban}

We consider random interferers on road-$x$ characterized by $\Phi_{\mathrm{x}}\sim\textrm{PPP}\left(p_{\mathrm{I}}\lambda_{\mathrm{x}},\mathcal{B}_{\mathrm{x}}\right)$.
We substitute the channel model of (\ref{eq: Channel - Suburban})
into (\ref{eq:  P(PPP) =00003D Theorem}), and since $\mathbf{x}_{\mathrm{rx}}=x_{\mathrm{rx}}\mathbf{e}_{\mathrm{x}}$
and $\mathbf{x}=x\mathbf{e}_{\mathrm{x}}$, the multidimensional integration
reduces to
\begin{align}
 & \mathcal{P_{\circ}}\left(\Phi_{\mathrm{x}}\right)=\exp\biggl(-\int_{\mathcal{B}_{\mathrm{x}}}\frac{p_{\mathrm{I}}\lambda_{\mathrm{x}}\ \mathrm{d}x}{\bigl(1+\bigl(\left|x_{\mathrm{rx}}-x\right|/\zeta_{\mathrm{s}}\bigr)^{\alpha}\bigr)}\biggr),\label{eqApp: A.1}
\end{align}
such that $\zeta_{\mathrm{s}}=\left(A_{\circ}\beta_{\mathrm{s}}^{\prime}\right)^{1/\alpha}=\beta^{1/\alpha}\left\Vert \mathbf{x}_{\mathrm{rx}}-\mathbf{x}_{\mathrm{tx}}\right\Vert $.
At present, to solve (\ref{eqApp: A.1}), two possible cases arise.

\subsubsection*{Case I \textendash{} RX is Inside Bounded Set $\mathcal{B}_{\mathrm{x}}$:
$\left\Vert \mathbf{x}_{\mathrm{rx}}\right\Vert \leq R_{\mathrm{x}}$}

Due to $\left|x_{\mathrm{rx}}-x\right|$, the above integral must
be split in two parts, namely
\begin{align}
 & \!\!\!\mathcal{P_{\circ}}\left(\Phi_{\mathrm{x}}\right)=\exp\biggl(-p_{\mathrm{I}}\lambda_{\mathrm{x}}\biggl\{\intop_{-R_{\mathrm{x}}}^{x_{\mathrm{rx}}}\frac{\mathrm{d}x}{\bigl(1+\bigl(\bigl(x_{\mathrm{rx}}-x\bigr)/\zeta_{\mathrm{s}}\bigr)^{\alpha}\bigr)}\nonumber \\
 & \;\;\;\;\;\;\;\;\;\;\;\;\;\;\;\;\;\;\;\;\;\;\;\;\;\;+\;\intop_{x_{\mathrm{rx}}}^{R_{\mathrm{x}}}\frac{\mathrm{d}x}{\bigl(1+\bigl(\bigl(x-x_{\mathrm{rx}}\bigr)/\zeta_{\mathrm{s}}\bigr)^{\alpha}\bigr)}\biggr\}\biggr)\mathrm{.}\label{eqApp: A.2}
\end{align}
If we let $u=\left(x_{\mathrm{rx}}-x\right)/\zeta_{\mathrm{s}}$ for
the first part, and $v=-u$ for the second, (\ref{eqApp: A.2}) becomes
\begin{align}
 & \mathcal{P_{\circ}}\left(\Phi_{\mathrm{x}}\right)=\label{eqApp: A.3}\\
 & \exp\biggl(\!-p_{\mathrm{I}}\lambda_{\mathrm{x}}\zeta_{\mathrm{s}}\biggl\{ g_{\circ}\Bigl(\alpha,\frac{\left(R_{\mathrm{x}}+x_{\mathrm{rx}}\right)}{\zeta_{\mathrm{s}}}\Bigr)\!+\!g_{\circ}\Bigl(\alpha,\frac{\left(R_{\mathrm{x}}-x_{\mathrm{rx}}\right)}{\zeta_{\mathrm{s}}}\Bigr)\!\biggr\}\!\!\biggr),\nonumber 
\end{align}
where the recurring function $g_{\circ}\left(\alpha,\vartheta\right)$
is defined below. Meanwhile, we should underscore that due to the
symmetry in (\ref{eqApp: A.3}), it is possible to replace $x_{\mathrm{rx}}$
by $\left\Vert \mathbf{x}_{\mathrm{rx}}\right\Vert $, while still
remaining compatible when $x_{\mathrm{rx}}<0$.
\begin{defn}
\label{MATH-Definition: G-Function}Let the function $g_{\circ}\left(\alpha,\vartheta\right)\!:\!\mathbb{R}^{+}\times\mathbb{R}_{0}^{+}\mapsto\mathbb{R}_{0}^{+}$
be dependent on Gauss's hypergeometric function $_{2}F_{1}\left(a,b;c;x\right)$
as follows
\begin{align}
 & \!\!\!g_{\circ}\!\left(\alpha,\vartheta\right)\triangleq\!\!\intop_{0}^{\vartheta}\!\!\frac{\mathrm{d}u}{\left(1+u^{\alpha}\right)}\!=\!\vartheta\thinspace{}_{2}F_{1}\!\!\left(\!\!1,\frac{1}{\alpha};\bigl(1+\frac{1}{\alpha}\bigr);-\vartheta^{\alpha}\!\!\right)\mathrm{.}\label{eqApp: A.4 =00005BG-Function=00005D}
\end{align}
\end{defn}
\noindent For some values of $\alpha>1$, the function in (\ref{eqApp: A.4 =00005BG-Function=00005D})
reverts to a simple form; for instance: $g_{\circ}\left(\alpha=2,\vartheta\right)=\arctan\left(\vartheta\right)$.

\subsubsection*{Case II \textendash{} RX is Outside Bounded Set $\mathcal{B}_{\mathrm{x}}$:
$\left\Vert \mathbf{x}_{\mathrm{rx}}\right\Vert >R_{\mathrm{x}}$}

The RX must be outside the region of H-PPP interferers on road-$x$;
therefore, we may consider $x_{\mathrm{rx}}<-R_{\mathrm{x}}$ or $x_{\mathrm{rx}}>R_{\mathrm{x}}$.
Due to symmetry, the final result will be identical. As it is more
evolved, here we opt to demonstrate the derivation with a RX positioned
on the negative axis. Thus, we replace $\left|x_{\mathrm{rx}}-x\right|$
by $\left(x-x_{\mathrm{rx}}\right)$ in (\ref{eqApp: A.1}), while
taking the integration over $\left|x\right|\leq R_{\mathrm{x}}$;
also, realizing that $-x_{\mathrm{rx}}=\left\Vert \mathbf{x}_{\mathrm{rx}}\right\Vert $,
we get
\begin{align}
 & \!\!\mathcal{P_{\circ}}\left(\Phi_{\mathrm{x}}\right)=\exp\biggl(-\intop_{-R_{\mathrm{x}}}^{R_{\mathrm{x}}}\frac{p_{\mathrm{I}}\lambda_{\mathrm{x}}\ \mathrm{d}x}{\bigl(1+\bigl(\bigl(x+\left\Vert \mathbf{x}_{\mathrm{rx}}\right\Vert \bigr)/\zeta_{\mathrm{s}}\bigr)^{\alpha}\bigr)}\biggr)\mathrm{.}\label{eqApp: A.5}
\end{align}
If we let $u=\left(x+\left\Vert \mathbf{x}_{\mathrm{rx}}\right\Vert \right)/\zeta_{\mathrm{s}}$,
the expression in (\ref{eqApp: A.5}) will then equal to (\ref{eqApp: A.6}),
where $g_{\circ}\left(\alpha,\vartheta\right)$ is defined in (\ref{eqApp: A.4 =00005BG-Function=00005D}).
\begin{align}
 & \mathcal{\!\!P_{\circ}}\left(\Phi_{\mathrm{x}}\right)=\exp\biggl(-p_{\mathrm{I}}\lambda_{\mathrm{x}}\zeta_{\mathrm{s}}\!\!\intop_{\left(\left\Vert \mathbf{x}_{\mathrm{rx}}\right\Vert -R_{\mathrm{x}}\right)/\zeta_{\mathrm{s}}}^{\left(\left\Vert \mathbf{x}_{\mathrm{rx}}\right\Vert +R_{\mathrm{x}}\right)/\zeta_{\mathrm{s}}}\!\!\frac{\mathrm{d}u}{\bigl(1+u^{\alpha}\bigr)}\biggr)=\label{eqApp: A.6}\\
 & \!\!\exp\biggl(\!-p_{\mathrm{I}}\lambda_{\mathrm{x}}\zeta_{\mathrm{s}}\biggl\{\!g_{\circ}\!\Bigl(\!\alpha,\!\frac{\left(\left\Vert \mathbf{x}_{\mathrm{rx}}\right\Vert \!+\!R_{\mathrm{x}}\right)}{\zeta_{\mathrm{s}}}\!\Bigr)\!-\!g_{\circ}\!\Bigl(\!\alpha,\!\frac{\left(\left\Vert \mathbf{x}_{\mathrm{rx}}\right\Vert \!-\!R_{\mathrm{x}}\right)}{\zeta_{\mathrm{s}}}\!\Bigr)\!\biggr\}\!\biggr)\nonumber 
\end{align}

\section{\label{AppB: Proposition2}Proof of Proposition \ref{MATH-Proposition: Suburban; Road-Y}:
$\mathcal{P_{\circ}}\left(\Phi_{\mathrm{y}}\right)$ for Suburban}

We analyze random interfering vehicles on road-$y$ modeled by $\Phi_{\mathrm{y}}\sim\textrm{PPP}\left(p_{\mathrm{I}}\lambda_{\mathrm{y}},\mathcal{B}_{\mathrm{y}}\right)$.
The RX is positioned at $\mathbf{x}_{\mathrm{rx}}=x_{\mathrm{rx}}\mathbf{e}_{\mathrm{x}}$
and the interferers are located at $\mathbf{x}=y\mathbf{e}_{\mathrm{y}}$.
After substituting these spatial parameters and the channel model
of (\ref{eq: Channel - Suburban}) into (\ref{eq:  P(PPP) =00003D Theorem}),
we find that
\begin{align}
 & \!\!\!\mathcal{P_{\circ}}\left(\Phi_{\mathrm{y}}\right)=\exp\biggl(-\int_{\mathcal{B}_{\mathrm{y}}}\frac{p_{\mathrm{I}}\lambda_{\mathrm{y}}\ \mathrm{d}y}{\bigl(1\!+\bigl(\bigl(y^{2}\!+\!\left\Vert \mathbf{x}_{\mathrm{rx}}\right\Vert ^{2}\bigr)\!/\!\zeta_{\mathrm{s}}^{2}\bigr)^{\alpha/2}\bigr)}\biggr)\mathrm{.}\label{eqApp: B.1}
\end{align}
The above could be simplified by assigning $u=\bigl(y^{2}+\left\Vert \mathbf{x}_{\mathrm{rx}}\right\Vert ^{2}\bigr)/\zeta_{\mathrm{s}}^{2}$,
and thus $\mathrm{d}y=\zeta_{\mathrm{s}}^{2}\mathrm{d}u/2y$, where
$y=\pm\thinspace\zeta_{\mathrm{s}}\sqrt{u-\left\Vert \mathbf{x}_{\mathrm{rx}}\right\Vert ^{2}/\zeta_{\mathrm{s}}^{2}}$.
Following some manipulations, we get
\begin{align}
 & \!\!\!\mathcal{P_{\circ}}\left(\Phi_{\mathrm{y}}\right)=\exp\biggl(-p_{\mathrm{I}}\lambda_{\mathrm{y}}\zeta_{\mathrm{s}}\!\!\!\!\intop_{\delta}^{R_{\mathrm{y}}^{2}/\zeta_{\mathrm{s}}^{2}+\delta}\!\!\!\!\frac{\mathrm{d}u}{\sqrt{u-\delta}\bigl(1+u^{\alpha/2}\bigr)}\biggr),\label{eqApp: B.2}
\end{align}
such that $\delta=\left\Vert \mathbf{x}_{\mathrm{rx}}\right\Vert ^{2}/\zeta_{\mathrm{s}}^{2}$.
The format of the integration within (\ref{eqApp: B.2}) can be defined
as follows.
\begin{defn}
\label{MATH-Definition: H-Function}Let the function $h_{\circ}\left(\alpha,\delta,\vartheta\right)\!:\!\mathbb{R}^{+}\times\mathbb{R}_{0}^{+}\times\mathbb{R}_{0}^{+}\mapsto\mathbb{R}_{0}^{+}$
be defined as follows
\begin{align}
 & h_{\circ}\left(\alpha,\delta,\vartheta\right)\triangleq\intop_{\delta}^{\vartheta+\delta}\frac{\mathrm{d}u}{\sqrt{u-\delta}\left(1+u^{\alpha/2}\right)}\thinspace\mathrm{.}\label{eqApp: B.3 =00005BH-Function=00005D}
\end{align}
\end{defn}
\noindent For values of $\alpha>1$, this integral can be solved efficiently
using standard numerical methods. However, in some cases, a tractable
expression for (\ref{eqApp: B.3 =00005BH-Function=00005D}) is available,
for instance
\begin{align}
 & \!\!\!\!\!\!\!\!\!\!\!h_{\circ}\left(\alpha,0,\vartheta\right)=2\sqrt{\vartheta}\thinspace{}_{2}F_{1}\left(1,\frac{1}{\alpha};\bigl(1+\frac{1}{\alpha}\bigr);-\vartheta^{\alpha/2}\right)\label{eqApp: B.4}\\
 & \!\!\!\!\!\!\!\!\!\!\!h_{\circ}\left(\alpha=2,\delta,\vartheta\right)=2\arctan\bigl(\sqrt{\vartheta/\left(1+\delta\right)}\thinspace\bigr)/\sqrt{1+\delta}\thinspace\mathrm{.}\label{eqAPP: B.5}
\end{align}

\section{\label{AppC: Proposition4}Proof of Proposition \ref{MATH-Proposition: Urban; Road-Y}:
$\mathcal{P_{\circ}}\left(\Phi_{\mathrm{y}}\right)$ for Urban}

We consider random interferers on road-$y$ modeled by $\Phi_{\mathrm{y}}\sim\textrm{PPP}\left(p_{\mathrm{I}}\lambda_{\mathrm{y}},\mathcal{B}_{\mathrm{y}}\right)$,
where the RX and interferes are accordingly located at $\mathbf{x}_{\mathrm{rx}}=x_{\mathrm{rx}}\mathbf{e}_{\mathrm{x}}$
and $\mathbf{x}=y\mathbf{e}_{\mathrm{y}}$. Taking into account these
parameters and substituting the propagation model for urban intersection
of (\ref{eq: Channel - Urban}) into (\ref{eq:  P(PPP) =00003D Theorem}),
we obtain
\begin{align}
 & \mathcal{P_{\circ}}\left(\Phi_{\mathrm{y}}\right)=\exp\biggl(-\int_{\mathcal{B}_{\mathrm{y}}}\frac{p_{\mathrm{I}}\lambda_{\mathrm{y}}\ \mathrm{d}y}{\bigl(1+1/\beta_{\mathrm{u}}^{\prime}\ell_{\mathrm{pl}}^{\mathrm{u}}\left(\mathbf{x},\mathbf{x}_{\mathrm{rx}}\right)\bigr)}\biggr)\mathrm{.}\label{eqAPP: C.1}
\end{align}
To solve (\ref{eqAPP: C.1}), two possible cases arise.

\subsubsection*{Case I \textendash{} RX is Near the Intersection Point: $\left\Vert \mathbf{x}_{\mathrm{rx}}\right\Vert \leq\triangle$}

When the RX is closer to the intersection, the WLOS Manhattan model
within (\ref{eq: Channel - Urban}) is the only relevant channel;
thus we get
\begin{align}
 & \!\!\!\mathcal{P_{\circ}}\left(\Phi_{\mathrm{y}}\right)=\exp\biggl(-\int_{\mathcal{B}_{\mathrm{y}}}\frac{p_{\mathrm{I}}\lambda_{\mathrm{y}}\ \mathrm{d}y}{\bigl(1+\bigl(\bigl(\left|y\right|+\left\Vert \mathbf{x}_{\mathrm{rx}}\right\Vert \bigr)/\zeta_{\mathrm{u}}\bigr)^{\alpha}\bigr)}\biggr),\label{eqApp: C.2}
\end{align}
where $\zeta_{\mathrm{u}}=\left(A_{\circ}\beta_{\mathrm{u}}^{\prime}\right)^{1/\alpha}$.
If we perform a change of variable to (\ref{eqApp: C.2}) with $u=\bigl(\left|y\right|+\left\Vert \mathbf{x}_{\mathrm{rx}}\right\Vert \bigr)/\zeta_{\mathrm{u}}$,
we obtain
\begin{align}
 & \mathcal{P_{\circ}}\left(\Phi_{\mathrm{y}}\right)=\label{eqApp: C.3}\\
 & \exp\biggl(-2p_{\mathrm{I}}\lambda_{\mathrm{y}}\zeta_{\mathrm{u}}\biggl\{ g_{\circ}\Bigl(\alpha,\frac{\bigl(R_{\mathrm{y}}+\left\Vert \mathbf{x}_{\mathrm{rx}}\right\Vert \bigr)}{\zeta_{\mathrm{u}}}\Bigr)-g_{\circ}\Bigl(\alpha,\frac{\left\Vert \mathbf{x}_{\mathrm{rx}}\right\Vert }{\zeta_{\mathrm{u}}}\Bigr)\!\biggr\}\!\biggr),\nonumber 
\end{align}
where $g_{\circ}\left(\alpha,\vartheta\right)$ is defined in (\ref{eqApp: A.4 =00005BG-Function=00005D}).

\subsubsection*{Case II \textendash{} RX is Away from the Intersection Point: $\left\Vert \mathbf{x}_{\mathrm{rx}}\right\Vert >\triangle$}

In this case, the WLOS Manhattan model within (\ref{eq: Channel - Urban})
is relevant for $\left\Vert \mathbf{x}\right\Vert \leq\triangle$,
and the NLOS VirtualSource11p model over $\triangle\!<\!\left\Vert \mathbf{x}\right\Vert \!\leq\!R_{\mathrm{y}}$.
Applying these models into (\ref{eqAPP: C.1}), we get
\begin{align}
 & \negthinspace\negthinspace\mathcal{P_{\circ}}\left(\Phi_{\mathrm{y}}\right)\!=\!\exp\Biggl(\!-2p_{\mathrm{I}}\lambda_{\mathrm{y}}\Biggl\{\intop_{0}^{\triangle}\!\frac{\mathrm{d}y}{\left(1+\left(\left(y+\left\Vert \mathbf{x}_{\mathrm{rx}}\right\Vert \right)/\zeta_{\mathrm{u}}\right)^{\alpha}\right)}\nonumber \\
 & \ \ \ \ \ \ \ \ \ \ \ \ \ \ \ \ \ \ \ \ \ \ \ +\ \intop_{\triangle}^{R_{\mathrm{y}}}\frac{\mathrm{d}y}{\left(1+\left(y\left\Vert \mathbf{x}_{\mathrm{rx}}\right\Vert /\zeta_{\mathrm{u}}^{\prime}\right)^{\alpha}\right)}\Biggr\}\Biggr),\label{eqApp: C.4}
\end{align}
where $\zeta_{\mathrm{u}}\negthinspace=\negthinspace\left(A_{\circ}\beta_{\mathrm{u}}^{\prime}\right)^{1/\alpha}$
and $\zeta_{\mathrm{u}}^{\prime}\negthinspace=\negthinspace\left(A_{\circ}^{\prime}\beta_{\mathrm{u}}^{\prime}\right)^{1/\alpha}\negthinspace=\negthinspace\zeta_{\mathrm{u}}\left(A_{\circ}^{\prime}/A_{\circ}\right)^{1/\alpha}$.
If we let $u=\left(y+\left\Vert \mathbf{x}_{\mathrm{rx}}\right\Vert \right)/\zeta_{\mathrm{u}}$
for the first integration in (\ref{eqApp: C.4}), and $v=y\left\Vert \mathbf{x}_{\mathrm{rx}}\right\Vert /\zeta_{\mathrm{u}}^{\prime}$
for the second, we get
\begin{align}
 & \mathcal{P_{\circ}}\left(\Phi_{\mathrm{y}}\right)=\exp\biggl(-2p_{\mathrm{I}}\lambda_{\mathrm{y}}\zeta_{\mathrm{u}}\biggl\{ g_{\circ}\Bigl(\alpha,\frac{\bigl(\triangle+\left\Vert \mathbf{x}_{\mathrm{rx}}\right\Vert \bigr)}{\zeta_{\mathrm{u}}}\Bigr)-\nonumber \\
 & g_{\circ}\Bigl(\alpha,\frac{\left\Vert \mathbf{x}_{\mathrm{rx}}\right\Vert }{\zeta_{\mathrm{u}}}\Bigr)\!+\!\frac{1}{\kappa}\biggl(g_{\circ}\Bigl(\alpha,\frac{\kappa R_{\mathrm{y}}}{\zeta_{\mathrm{u}}}\Bigr)-g_{\circ}\Bigl(\alpha,\frac{\kappa\triangle}{\zeta_{\mathrm{u}}}\Bigr)\biggr)\!\biggr\}\!\biggr),\label{eqApp: C.5}
\end{align}
where $\kappa=\left(A_{\circ}/A_{\circ}^{\prime}\right)^{1/\alpha}\left\Vert \mathbf{x}_{\mathrm{rx}}\right\Vert $
and $g_{\circ}\left(\alpha,\vartheta\right)$ is defined in (\ref{eqApp: A.4 =00005BG-Function=00005D}).

\section*{Acknowledgments}

The authors are also grateful to Prof. Martin Haenggi (University
of Notre Dame, IN, USA) for discussions regarding the meta distribution.

\bibliographystyle{IEEEtran}
\bibliography{refJP17}

\end{document}